\documentclass[twocolumn]{aastex62}

\usepackage[utf8]{inputenc}
\usepackage{amsmath,amsxtra,amssymb,latexsym, amscd,amsthm}
\usepackage[mathscr]{eucal}
\usepackage{mathrsfs}
\usepackage{makeidx}
\usepackage{soul}
\setstcolor{red}

\newcommand{\mrm}[1]{\mathrm{#1}}
\newcommand{\mbf}[1]{\mathbf{#1}}

\begin{document}

\title{Meso-scale Instability Triggered by Dust Feedback in Dusty Rings: Origin and Observational Implications}

\correspondingauthor{Pinghui Huang}
\email{phhuang10@gmail.com, phhuang@pmo.ac.cn}

\author{Pinghui Huang}
\affiliation{CAS Key Laboratory of Planetary Sciences, Purple Mountain Observatory, Chinese Academy of Sciences, Nanjing 210008, China}
\affiliation{Institute for Advanced Study, Tsinghua University, Beijing 100084, China}
\affiliation{University of Chinese Academy of Sciences, Beijing 100049, China}
\affiliation{Theoretical Division, Los Alamos National Laboratory, Los Alamos, NM 87545, USA}
\affiliation{Department of Physics \& Astronomy, Rice University, 6100 Main Street, Houston, TX 77005, USA}
\author{Hui Li}
\affiliation{Theoretical Division, Los Alamos National Laboratory, Los Alamos, NM 87545, USA}
\author{Andrea Isella}
\affiliation{Department of Physics \& Astronomy, Rice University, 6100 Main Street, Houston, TX 77005, USA}
\author{Ryan Miranda}
\affiliation{Institute for Advanced Study, Princeton, NJ 08540, USA}
\author{Shengtai Li}
\affiliation{Theoretical Division, Los Alamos National Laboratory, Los Alamos, NM 87545, USA}
\author{Jianghui Ji}
\affiliation{CAS Key Laboratory of Planetary Sciences, Purple Mountain Observatory, Chinese Academy of Sciences, Nanjing 210008, China}
\affiliation{CAS Center for Excellence in Comparative Planetology, Hefei 230026, China}

\begin{abstract}
High spatial resolution observations of protoplanetary disks (PPDs) by ALMA have revealed many details that are providing interesting constraints on the disk physics as well as dust dynamics, both of which are essential for understanding planet formation. We carry out high-resolution, 2D global hydrodynamic simulations, including the effects of dust feedback, to study the stability of dusty rings. When the ring edges are relatively sharp and the dust surface density becomes comparable to the gas surface density, we find that dust feedback enhances the radial gradients of both the azimuthal velocity profile and the potential vorticity profile at the ring edges. This eventually leads to instabilities on meso-scales (spatial scales of several disk scale heights), causing dusty rings to be populated with many compact regions with highly concentrated dust densities on meso-scales. We also produce synthetic dust emission images using our simulation results and discuss the comparison between simulations and observations.

\end{abstract}

\keywords{instabilities --- hydrodynamics --- protoplanetary disks --- submillimeter: planetary systems}

\section{Introduction}\label{sec:intro}
Over the past few years, observations obtained with the Atacama Large Millimeter Array (ALMA) have revealed small scale structures in the distribution of (sub)millimeter dust grains around many circumstellar disks surrounding nearby young stars. The majority of the millimeter bright disks observed so far feature multiple rings and gaps characterized by radii spanning from a few to a few hundreds of astronomical units (au) \citep{almapartnetship2015, andrews2016ringed, isella2016ringed, andrews2018science, huang2018co, fedele2018alma, perez2018first, andrews2018disk, huang2018disk2, guzman2018disk, dullemond2018disk, isella2018disk}, while a smaller fraction of objects show prominent dust crescents \citep{fujiwara2006asymmetric,isella2010millimeter,isella2013azimuthal,van2013major,perez2014large,van2016vortices,boehler2017close,benisty2018shadows,cazzoletti2018evidence,perez2018disk,boehler2018complex} and spiral arms \citep{grady2012spiral,benisty2015asymmetric,akiyama2016spiral,boehler2018complex,dong2018eccentric,huang2018disk3,uyama2018subaru}.

The origin of these structures is debated but it is common understanding that they might carry key information about the formation of planets. One hypothesis is that these features originate from the interaction between newly formed planets and the circumstellar material \citep{ou2007disk,fung2014empty,zhu2014dust,dipierro2015planet,dong2015observational,jin2016modeling,liu2018new,van2018rings,jin2019new}. Numerical simulations indicate that azimuthally symmetric rings might result from the gravitational torque exerted by planets with masses below that of Jupiter \citep{bryden1999tidally,bryden2000interaction,bae2017formation,dong2017multiple,dong2018multiple}, while crescents might be caused by the onset of the Rossby Wave Instability at the unstable gap edge of more massive planets \citep[RWI,][]{lovelace1999rossby, li2000rossby, li2001rossby, li2005potential, fu2014long, lovelace2014rossby, zhu2014particle, huang2018identifying}. In an alternative scenario, rings and crescents form from radial variations in the disk viscosity, and consequently, in the gas radial drift velocity, which, in turn, leads to the accumulation of dust and gas in circular rings \citep{regaly2011possible,regaly2013trapping,flock2015gaps,huang2019observability}. If the viscosity variation is steep, the rings might become Rossby Wave unstable and generate gas vortices capable of trapping dust particles \cite[][hereafter M17]{miranda2017long}. A third hypothesis is that rings form at the snow line of volatile molecules \citep{zhang2015evidence,zhang2016commonality,yen2016gas}. However this scenario does not seem to be consistent with the rather random distribution of rings radii \citep{huang2018co, long2018gaps}.

Regardless of their physical origin, it has been shown that dusty rings and crescents are places where the density of sub-millimeter dust particles is enhanced compared to that of the circumstellar gas \citep{isella2016ringed, boehler2017close, boehler2018complex, dullemond2018dust}. For example, in the case of HD~142527 \citep{boehler2017close}, the dust-to-gas ratio within the prominent crescent that characterizes this disk is so large ($\sim$1.5) that we expect the dynamics of the system to be influenced by the dust component. When this happens, instabilities such as the streaming instability \citep{youdin2005streaming,youdin2007protoplanetary,johansen2007protoplanetary} are likely to grow and lead to the formation of dense dust clumps, and, perhaps, planetary embryos. Other systems featuring crescents and rings show more moderate variations in the dust-to-gas ratio, but they might nevertheless influence the overall dynamics of the disk. For example, the dust-to-gas ratio in the crescents observed around MWC~758 reaches a maximum of 0.2, while in the multiple ring systems observed around HL~Tau and HD~163296, the dust-to-gas ratio is constrained to be as high as about 0.05 \citep{jin2016modeling,isella2016ringed}, i.e., 5 times higher than the typical dust-to-gas ratio in the interstellar medium.

The effect of a high dust-to-gas ratio on the evolution of a vortex generated by RWI induced by the planet-disk interaction has been studied by \cite{fu2014long}, who found that dust feedback might lead to the destruction of the vortex itself, and, consequently, to the corresponding dust crescent, in about $10^3$-$10^4$ orbits of the planet responsible for the perturbation. For example, a dust crescent generated by a massive planet orbiting at 10 au around a solar mass star might live for less than about $3\times10^5$ yr. Alternatively, M17 found that a crescent caused by a sharp radial decrease of the gas viscosity might have a life time comparable to that of the circumstellar disk itself, thanks to the continuous replenishing of gas accreting inward from the outer disk regions. Despite the differences in the predicted vortex lifetime, these studies indicate that a high dust-to-gas ratio might lead to the fragmentation of dusty rings into dense clumps.

In this work, we extend the analysis of the effect of dust feedback on meso-scale disk structures by investigating the evolution of dusty rings. Similar to M17, we generate gas and dusty rings by imposing radial variations of the disk viscosity with a low viscosity region mimicking the dead-zone. Different from M17, however, we set the viscosity gradient so that the resulting radial density profile after a few thousand orbits is not steep enough to trigger RWI for the gas component. The presence of the gas surface density ring causes the dust to continue accumulating in this region. Ultimately, as the dust to gas ratio approaches unity in this ring, we find that instabilities get excited due to the dust feedback on the gas and, in certain cases, a large number of dust clumps are formed. In this work, we investigate the conditions for such an instability to develop, study the properties and evolution of the resulting clumps, and discuss the possibility of observing such structures using ALMA.

The outline of this paper is as follows. In Section~\ref{sec:setup}, we describe the setup of hydrodynamic model and the radiative transfer simulations. In Section~\ref{sec:hydro_results}, we present the hydrodynamical results of dust and gas, and discuss the dependence of instabilities on the adopted initial conditions.In Section~\ref{sec:images_results}, we present the synthetic ALMA images of our simulations. Discussion of the results in the framework of current and future observations are presented in Section~\ref{sec:discussion}, while the conclusions are summarized in Section~\ref{sec:summary}.

\section{Numerical Setup}\label{sec:setup}
\subsection{Hydrodynamic Model}
We calculate the temporal evolution of the circumstellar gas and dust using the multi-fluid hydrodynamic code LA-COMPASS \citep{li2005potential,li2008type,fu2014effects}. We assume a stellar mass $M_\star = \mrm{1\;M_\sun}$ and a normalization radius $R_0 = \mrm{50\;au}$, so that 1 orbit corresponds to a time interval of 353.5 years. The inner and outer boundaries of the numerical simulation are set to be $0.4\times R_0 = \mrm{20\;au}$ and $12\times R_0 = \mrm{600\;au}$, respectively. Initially, the gas disk surface density $\Sigma_\mrm{g}$ follows the viscous disk profile \citep{andrews2010protoplanetary}:
\begin{equation}
\label{eq:density}
\Sigma_\mrm{g} = 0.377 \left(\frac{R}{R_0}\right)^{-\gamma} \exp{\left(-\left(\frac{R}{R_\mrm{c}}\right)^{2-\gamma}\right)}
\end{equation}
where $\gamma = 0.8$, $R_\mrm{c} = 4\times R_0 = 200\;\mrm{au}$. The initial gas surface density at $R_0$ is $\Sigma_0 = 0.312\;\mrm{g\;cm^{-2}}$ and the total gas mass is 3 Jupiter masses (M$_\mrm{J}$). At the beginning of the simulation, the dust surface density ($\Sigma_\mrm{d}$) also follows Equation~\ref{eq:density} with a dust-to-gas ratio of 0.05, corresponding to a total dust mass of about 45 Earth masses (M$_\mrm{\earth}$). The adopted initial dust-to-gas ratio is larger than the typical value of 0.01 but is consistent with the observations of disks in Lupus \citep{ansdell2016alma}. Furthermore, as discussed in Section~\ref{sec:discussion}, we believe that the choice of the initial dust-to-gas ratio has no strong effects on the results of our analysis.

We calculate the disk evolution by neglecting the disk self-gravity and any effect due to magnetic fields. We treat dust and gas as fluids controlled by the following continuity and momentum equations \citep{fu2014effects}
\begin{equation}
\frac{\partial \Sigma_\mrm{d}}{\partial t} +\nabla \cdot \left(\Sigma_\mrm{d} \mbf{V}_\mrm{d}\right) = \nabla \cdot \left(\Sigma_\mrm{g} D_\mrm{d} \nabla \left(\frac{\Sigma_\mrm{d}}{\Sigma_\mrm{g}}\right)\right),
\label{eq:dust_con}
\end{equation}

\begin{equation}
\frac{\partial \Sigma_\mrm{g}}{\partial t} + \nabla \cdot \left(\Sigma_\mrm{g} \mbf{V}_\mrm{g}\right) =0,
\label{eq:gas_con}
\end{equation}

\begin{equation}
\frac{1}{\Sigma_\mrm{d}}\frac{\partial \Sigma_\mrm{d}\mbf{V}_\mrm{d}}{\partial t} +\frac{\nabla\left(\mbf{V}_\mrm{d}\cdot\Sigma_\mrm{d} \mbf{V}_\mrm{d}\right)}{\Sigma_\mrm{d}}=-\nabla\Psi_\mrm{G}+\mbf{F}_\mrm{drag},
\label{eq:dust_mom}
\end{equation}

\begin{equation}
\begin{aligned}
    \frac{1}{\Sigma_\mrm{g}}\frac{\partial \Sigma_\mrm{g}\mbf{V}_\mrm{g}}{\partial t} + \frac{\nabla\left(\mbf{V}_\mrm{g}\cdot\Sigma_\mrm{g} \mbf{V}_\mrm{g}\right)}{\Sigma_\mrm{g}}=&\\
-\nabla\Psi_\mrm{G}-\frac{\nabla P_\mrm{g}}{\Sigma_\mrm{g}}+&\mbf{F}_{\nu}-\frac{\Sigma_\mrm{d}}{\Sigma_\mrm{g}} \mbf{F}_\mrm{drag},
\end{aligned}
\label{eq:gas_mom}
\end{equation}
where $D_\mrm{d}$ is the dust mass diffusivity, $\Psi_\mrm{G}$ is the stellar gravitational potential, and $\mbf{V}_\mrm{d,g}$ are the dust and gas velocities. The viscous force is expressed by $\mbf{F}_\nu$, while $\mbf{F}_\mrm{drag}$ is the gas drag given by
\begin{equation}
\mbf{F}_\mrm{drag} = \frac{\Omega_\mrm{K}}{St} \left(\mbf{V}_\mrm{g} - \mbf{V}_\mrm{d}\right),
\label{eq:drag}
\end{equation}
where $St$ is the Stokes number of the dust particles in Epstein regime expressed as \citep{takeuchi2002radial}:
\begin{equation}
\label{eq:stokes}
    St = \frac{\pi \rho_\mrm{p} a_\mrm{p}}{2 \Sigma_\mrm{g}}~~,
\end{equation}
where $\rho_\mrm{p}$ and $a_\mrm{p}$ are the internal density and radius of dust particles, respectively. The last term of the right hand-side of Equation~\ref{eq:gas_mom} is the dust feedback term. The dust feedback term usually is weak in interstellar mediums because of the low dust-to-gas ratio. However, in protoplanetary disks, dust will be settled around mid-plane until it is balanced by turbulence stirring then form thin dust sub-layers~\citep{chiang2010forming,testi2014dust} and be trapped in pressure bumps to form dusty rings. Or gas can be removed by photoevaporation~\citep{alexander2006photoevaporationI,alexander2006photoevaporationII} and MHD disk wind~\citep{blandford1982hydromagnetic,bai2013wind1,bai2013wind2,bai2016magneto}. All these effects can increase the dust-to-gas ratio in protoplanetary disks then enhance the role of dust feedback.

In our analysis, we assume a single dust species characterized by $\rho_\mrm{p} = 2.0\;\mrm{g\;cm^{-3}}$ and $a_\mrm{p} = 0.25\;\mrm{mm}$. The initial $St$ at $R_0$ is about 0.25. The justification of this choice will be discussed in Section~\ref{subsubsec:stokes}. We evolve the system assuming a vertically isothermal equation of state, $P_\mrm{g} = c_\mrm{s}^2 \Sigma_\mrm{g}$, where $P_\mrm{g}$ and $c_\mrm{s}$ are the vertically integrated pressure and sound speed of gas. The radial profile of the sound speed is described as $c_\mrm{s}(R) = c_\mrm{s,0}(R/R_0)^{-1/4}$. The assumed sound speed corresponds to a disk aspect ratio $H_0/R_0 = 0.05$, where $H_0$ is the pressure scale height of disk ($H = c_\mrm{s}/\Omega_\mrm{K}$, $\Omega_\mrm{K}$ is the Keplerian angular frequency). The disk temperature profile corresponding to the adopted sound speed is $T\left(R\right) = 12.4\left(R/R_0\right)^{-1/2}\;\mrm{K}$ and it agrees with the disk temperature calculated using the radiative transfer simulations discussed in the next section.

In our hydrodynamic simulations, the kinematic viscosity is parameterized as $\nu = \alpha c_\mrm{s} H$, where $\alpha$ is the dimensionless viscous parameter \citep{shakura1973black,pringle1981accretion}. We further assume that $\alpha$ depends on the radius following a prescription designed to mimic the presence of a low viscosity region in the disk similar to that predicted by magnetorotational instability \citep[MRI,][]{balbus1998instability} dead-zone models \citep{gammie1996layered,armitage2011dynamics}. In this paper, we consider the Ohmic dead-zone viscosity transition model, which ignore the Hall effect~\citep{bai2014hall1,bai2014hall2} and the ambipolar diffusion~\citep{bai2011ambipolar}. Following \cite{regaly2011possible}, we assume that the radial profile of $\alpha$ in presence of a dead-zone is parameterized as
\begin{equation}
\label{eq:alpha}
    \alpha\left(R\right) = \alpha_0 - \left(\frac{\alpha_0-\alpha_{\mrm{DZ}}}{2}\right)\left( 1-\tanh\left(\frac{R-R_{\mrm{DZ}}}{\Delta_{\mrm{DZ}}}\right)\right)
\end{equation}
where $\alpha_0 = 10^{-2}$ and $\alpha_{\mrm{DZ}} = 10^{-5}$ are the viscosity parameters outside and within the dead-zone, respectively.

The transition between low and high viscosity happens at $R_\mrm{DZ} = 1.5 R_0 = 75$ au, and the width of the transition zone is controlled by the parameter $\Delta_{\mrm{DZ}}$. The transition from high to low viscosity leads to a corresponding formation of a gas over-density between $R_0$ and $R_\mrm{DZ}$. This over density leads to a local maximum in the gas pressure that is capable of forming a dusty ring by trapping dust particles as described in \cite{weidenschilling1977aerodynamics}.

\begin{figure}[!t]
\centering
\includegraphics[width=0.50\textwidth]{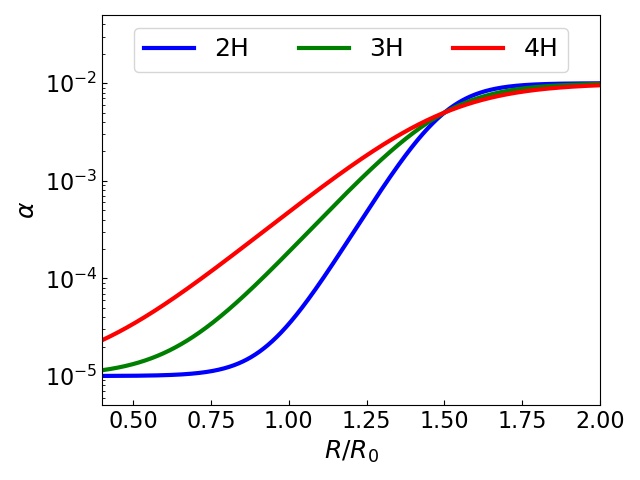}
\caption{Radial profile of the viscosity parameter $\alpha$ for the three models studied in this paper. The blue, green, and red lines correspond to models in which the transition between the low viscosity inner disk and the high viscosity outer disk $\Delta_{\mrm{DZ}}$ is equal to $2H$, $3H$ and $4H$, respectively, where $H$ is the pressure scale height of the disk.}
\label{fig:viscosity}
\end{figure}

Numerical simulations have shown that if $\Delta_{\mrm{DZ}} < 2H$, this ring becomes Rossby wave unstable and develop a large-scale vortex \citep[M17,][]{lyra2009planet, regaly2011possible}. Since the focus of this paper is on azimuthally symmetric rings, we avoid the formation of a large scale vortex by adopting $\Delta_{\mrm{DZ}} \geq 2H$. More precisely, we explore three cases in which $\Delta_{\mrm{DZ}}$ is equal to $2H, 3H$ and $4H$. The corresponding radial profiles of $\alpha$ are shown in Figure~\ref{fig:viscosity}.

The hydrodynamic simulations are performed on a 2D numerical grid characterized by 6144 cells both in the radial and azimuthal directions, and open boundary conditions. At $R_0$, the disk pressure scale height is resolved by about 90 cells. In order to speed up the hydrodynamic simulations, we run the first 1000 orbits of the $3H$ model and the first 1500 orbits of the $4H$ model using the 1D version of LA-COMPASS, and switch to 2D when the dust to gas ratio in the dusty rings approaches unity. A low level of noise is added to the 2D flow to seed non-axisymmetric instabilities. Performing the 1D simulation does not affect our results since no instabilities develop at the beginning of the simulation in these two models.

\subsection{Radiative Transfer and Synthetic Images}\label{subsec:radiative_images}
The outputs of LA-COMPASS 2D hydrodynamic simulations are post-processed using the radiative transfer code RADMC-3D\footnote{\url{http://www.ita.uni-heidelberg.de/~dullemond/software/radmc-3d/}}~\citep{dullemond2012radmc} and the Common Astronomy Software Applications (CASA)\footnote{\url{https://casa.nrao.edu/}}~\citep{mcmullin2007casa} to generate synthetic images of the dust continuum emission at the wavelength of 1.3 mm (230 GHz, ALMA Band 6). The adopted procedure is similar to that described in \cite{jin2016modeling} and \cite{liu2018ngvla}. In brief, we convert the 2D gas surface density ($\Sigma_\mrm{g}$) into a 3D density distributions ($\rho_\mrm{g}$) assuming hydrostatic equilibrium for a vertically isothermal disk such that $\rho_\mrm{g}\left(R,\Phi,Z\right) = \frac{\Sigma_\mrm{g}\left(R,\Phi\right)}{\sqrt{2\pi}H\left(R\right)}\exp\left({\frac{-Z^2}{2H\left(R\right)^2}}\right)$ \cite[][Chapter 5]{frank2002accretion}.

We then assume that the disk temperature is controlled by sub-micron size dust particles which are well coupled to the gas, and therefore follow the gas density distribution scaled for a constant dust-to-gas ratio equal to 0.01. The disk temperature is calculated using RADMC-3D adopting a dust opacity proper of interstellar grains made of astronomical silicates \citep{weingartner2001dust}, organic carbonates \citep{zubko1998size}, and water ice with fractional abundances as in \cite{pollack1994composition} and a MNR grain size distribution ($n(a) \propto a^{-3.5}$) with a minimum grain size of 0.5 $\mu$m and a maximum grain size of 10 mm. The calculation of dust opacity is discussed in \cite{isella2009structure}. The disk temperature of mid-plane at $R_0$ calculated by RADMC-3D and that corresponding to the sound speed adopted in the hydrodynamic simulations differ about 5\%, and we believe that the difference in temperature is sufficiently small that it should not affect our results.

After the temperature has been calculated, we use the ray-tracing method of RADMC-3D to generate synthetic images of the dust continuum emission at the wavelength of 1.3 mm. To do this, we convert the surface density of 0.25 mm grains generated by the hydrodynamic simulations into a 3D density distribution as above, but we now assume that the pressure scale height of the dust disk is 10\% of that of the gas \citep{pinte2015dust, isella2016ringed}. Note however that, since we present face on ($i = 0^\circ$) images of the dust emission, variations in the adopted dust scale height will not significantly affect our results. The dust opacity at (sub-)millimeter wavelength is assumed to be $2.3\;\mrm{cm^2\;g^{-1}}\times(\nu/230\;\mrm{GHz})^{0.4}$\citep{beckwith1990survey} and the disk is assumed to be at a distance of 140 pc from Earth.

Finally, synthetic disk images are processed using the ``simobserve'' task of CASA to simulate ALMA observation that properly account for observational noise. We adopt an array configuration that deliver a synthetic beam FWHM of $0.039\arcsec \times 0.036\arcsec$ when a Briggs robust parameter of 0.5 is adopted in the image deconvolution. We integrate for 3 hours under typical weather conditions, resulting in an RMS noise of 6.1 $\mu$Jy beam$^{-1}$.

\section{Hydrodynamical simulation results} \label{sec:hydro_results}
\subsection{General properties}

\begin{figure*}[!t]
\centering
\includegraphics[width=1.0\textwidth]{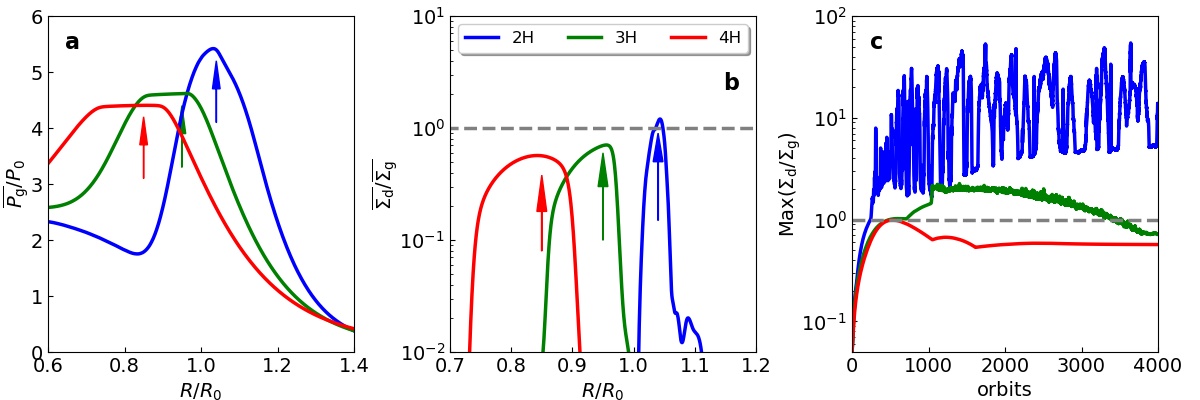}
\caption{Panel a: Azimuthally averaged radial profile of the gas pressure at orbit 4000 normalized
by $P_0 = c_\mrm{s,0}^2 \Sigma_0$. The blue, green, and red curves correspond to the $2H$, $3H$, and $4H$, models. The vertical arrows mark the location of the local pressure maxima induces by the radial viscosity variations. Panel b: Azimuthally averaged radial profile of the dust-to-gas ratio at orbit 4000. The gray dashed line indicate a dust-to-gas ratio equal to unity. Dense dusty rings form around the location of the gas pressure maxima. Panel c: Maximum of the dust-to-gas ratio as a function of time.
}
\label{fig:hydro_summary}
\end{figure*}

\begin{figure*}[!t]
\centering
\includegraphics[width=1.0\textwidth]{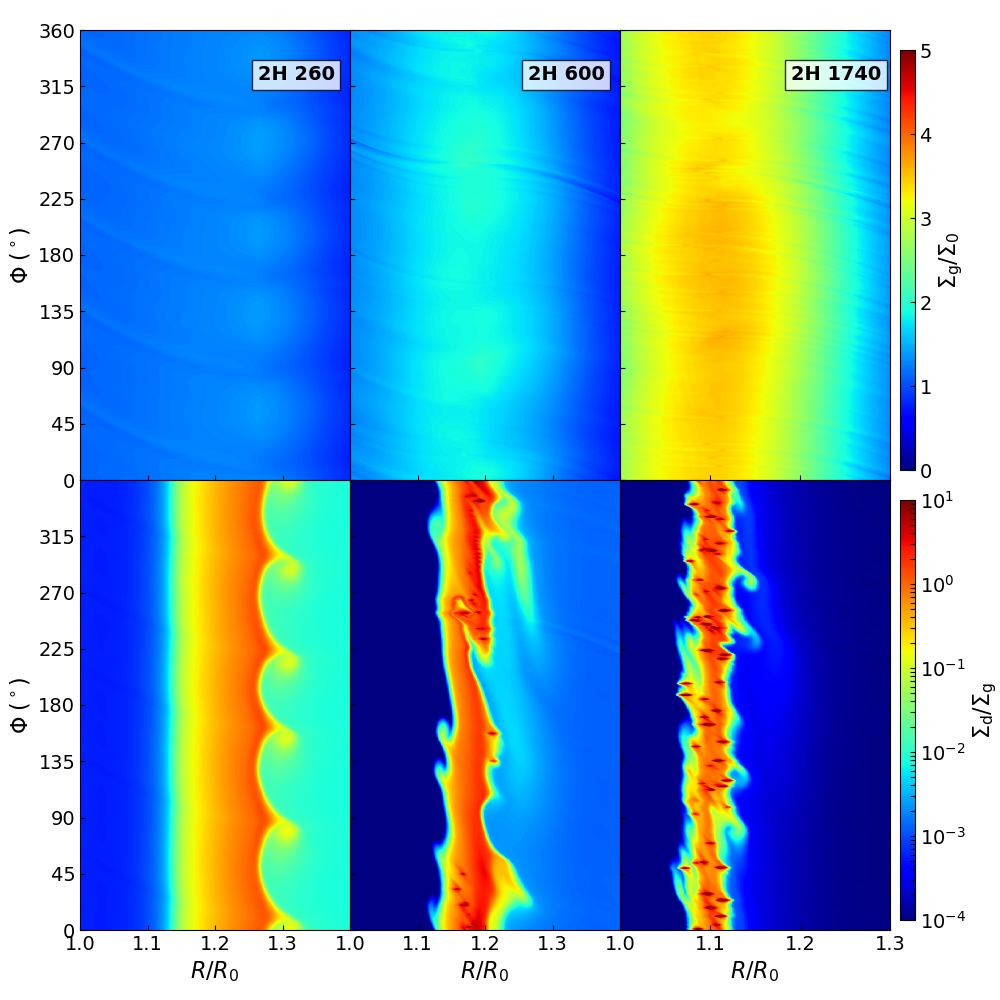}
\caption{Top row: polar maps of the gas surface density for the model $2H$ at orbit 260 (left), 600 (center), and 1740 (right) normalized by initial gas surface density $\Sigma_0$. Bottom row: snapshots of dust-to-gas ratio at the same orbits. At orbit 260, an instability occurs at the outer edge of dusty ring. At orbit 600, small dusty vortices (dust clumps) form closed to the outer edge of the dusty ring while its inner edge becomes unstable. At orbit 1740, more than 50 vortices characterized by dust-to-gas ratio above unity are present along the dusty ring. We will discuss the properties of dust clumps in Section~\ref{subsubsec:GI} and Figure~\ref{fig:hist_clumps}. }
\label{fig:density_2H_1}
\end{figure*}

\begin{figure*}[!t]
\centering
\includegraphics[width=1.0\textwidth]{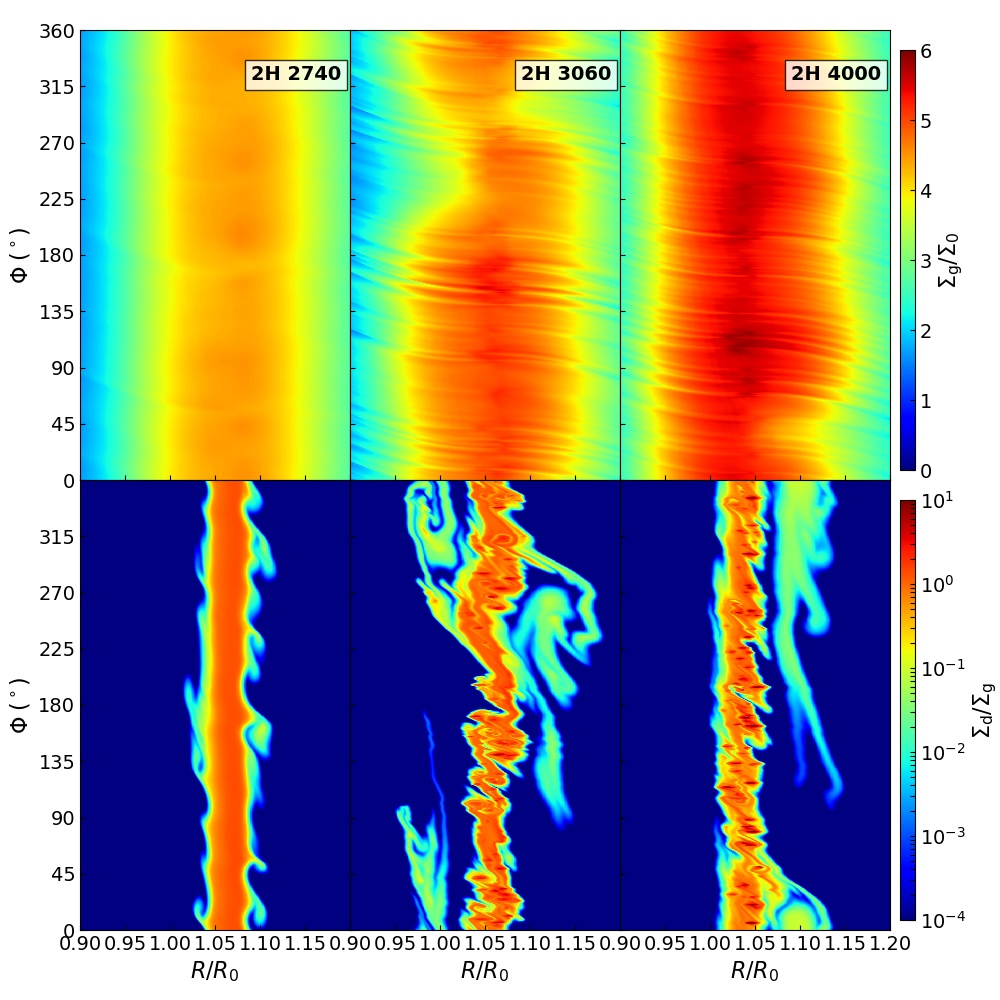}
\caption{Same as Figure~\ref{fig:density_2H_1}, but for orbit 2740, 3060 and 4000. At orbit 2740, the edges of the dusty ring show billow-shaped features but no dense clump are present. At orbit 3060 and 4000, the dusty ring shows both strong instabilities and several dense clumps.}
\label{fig:density_2H_2}
\end{figure*}

\begin{figure*}[!t]
\centering
\includegraphics[width=1.0\textwidth]{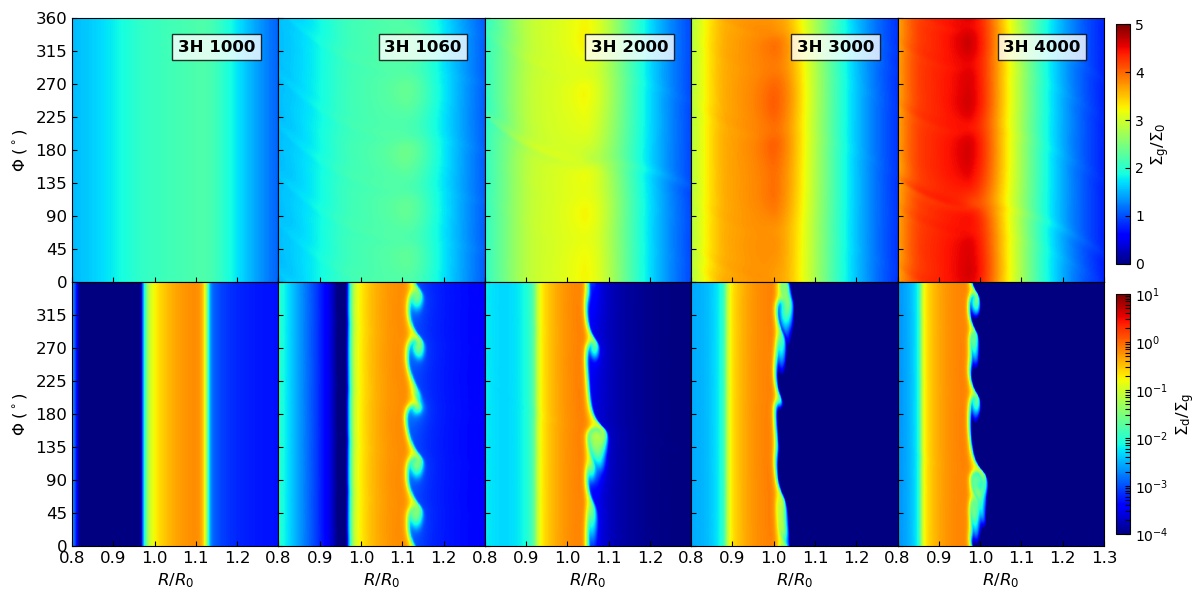}
\caption{Same as Figure~\ref{fig:density_2H_1}, but for the $3H$ model. Panels from left to right correspond to orbit 1000, 1060, 2000, 3000 and 4000. Billow-shaped features form at the outer edge of the dusty ring around orbit 1060. Differently from the $2H$ case, dense dust clumps never form in this model.}
\label{fig:density_3H}
\end{figure*}

\begin{figure*}[!t]
\centering
\includegraphics[width=1.0\textwidth]{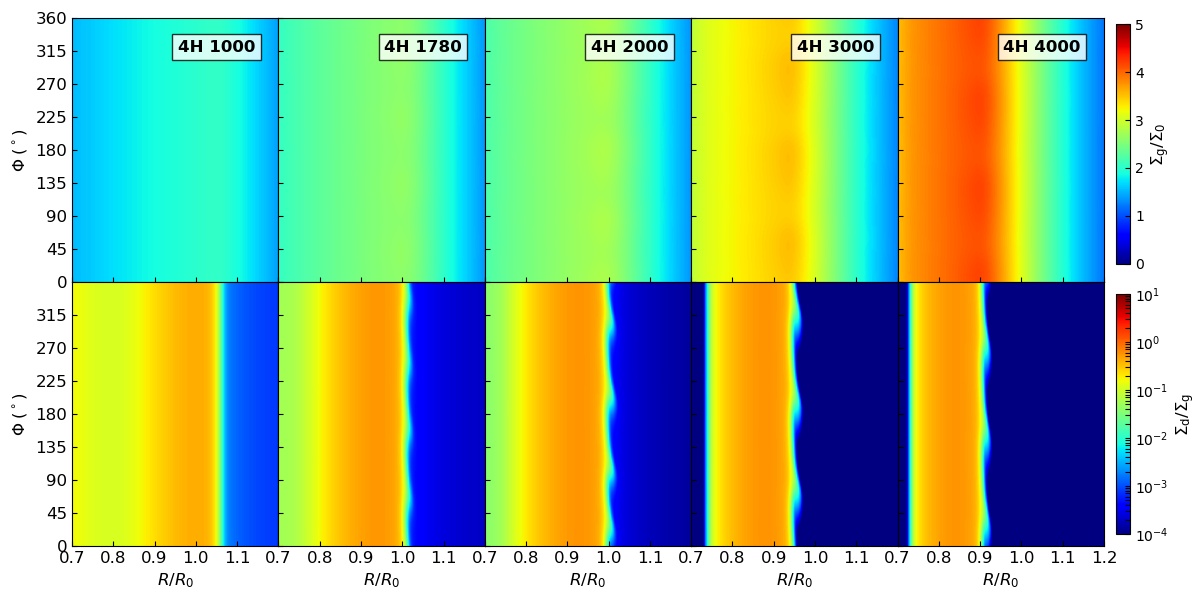}
\caption{
Same as Figure~\ref{fig:density_2H_1}, but for the $4H$ model. Panels from left to right correspond to orbit 1000, 1780, 2000, 3000 and 4000. Weak billow-shaped features form at the outer edge of the dusty ring around orbit 1780. As in the $3H$ case, no dense dust clumps form in this model.}
\label{fig:density_4H}
\end{figure*}

\begin{figure*}[!t]
\centering
\includegraphics[width=1.0\textwidth]{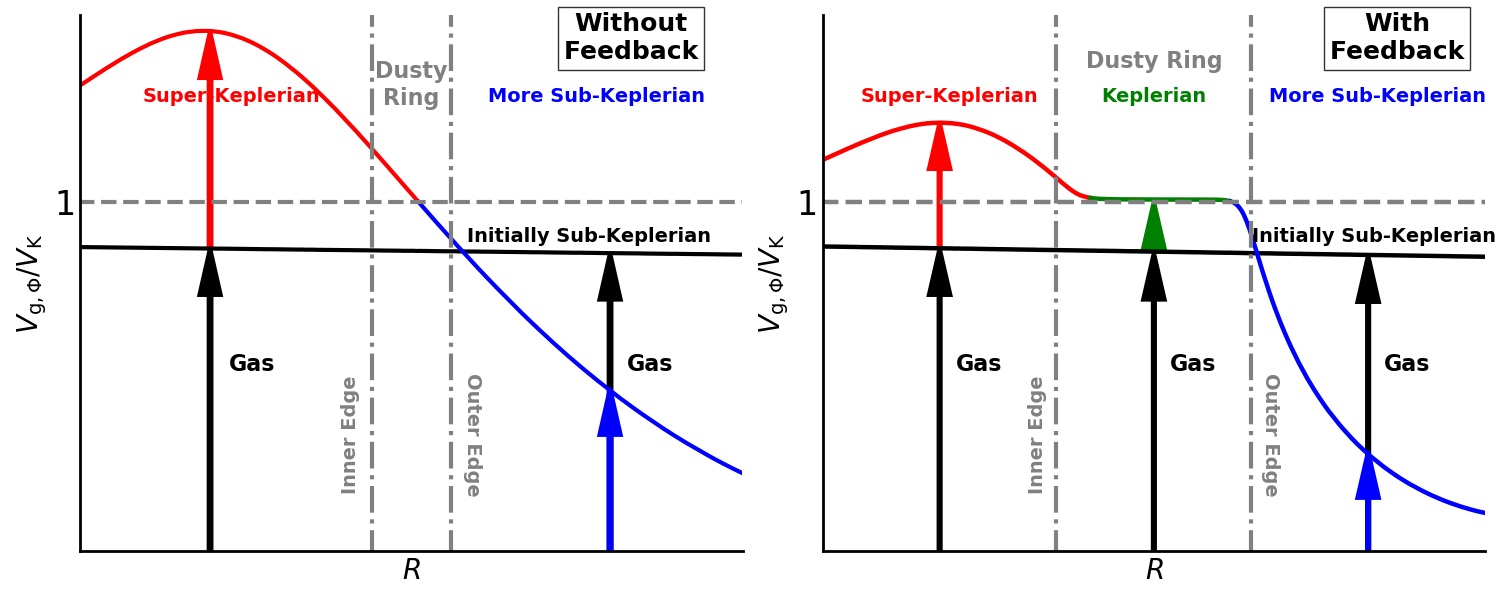}
\caption{Schematic diagram of the evolution of gas velocity around the position of the gas pressure maximum without (left) and with dust feedback (right). The x axis indicates the orbital radius normalized to the radius of the dusty ring forming due to the converging migration of solid particles toward the gas pressure maximum. The y axis indicates the azimuthal velocity of gas normalized to the local Keplerian velocity (grey dashed lines). At the beginning of our simulation, the gas pressure decreases with the orbital radius. This results in a slightly sub-Keplerian gas rotational velocity (black). As the pressure bump develops, dust is trapped in a radially narrow dusty ring (grey dashed-dotted lines). The gas inside and outside the dusty ring rotates slightly faster (red) and slower (blue) than Keplerian, respectively. Without dust feedback, the velocity of the gas across the dusty ring is not affected by the increasing dust-to-gas ratio. When the dust feedback is included, solid particles trapped inside the dusty ring push the gas to move at Keplerian velocity (green). The steep velocity gradient at the outer edge of the ring leads to onset of the instability. This instability at the inner edge of dusty rings in $2H$ happens later than that at outer edge, because it needs to take some time to produce strong enough gradient at the inner edge.}
\label{fig:origin}
\end{figure*}

\begin{figure*}[!t]
\centering
\includegraphics[width=1.0\textwidth]{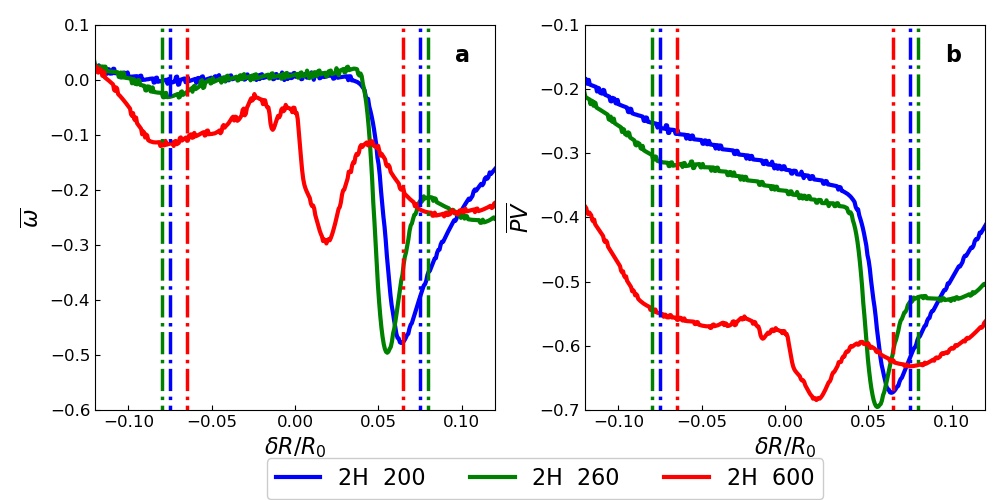}
\caption{Panel a: Azimuthally averaged vorticity of gas. Panel b: Azimuthally averaged potential vorticity. Blue, green, and red solid curves correspond to time frames in which instability has not developed yet (orbit 200), in a linear regime (orbit 260), and in a non linear regime (orbit 600), respectively. The vorticity and potential vorticity are all normalized by code units and Keplerian components are subtracted. The x-axis indicates radial distances from the center of dusty ring divided normalized by ring radius. The vertical dashed-dotted lines mark the inner and outer edges of dusty rings at different orbits.}
\label{fig:Vorticity_PV_2H}
\end{figure*}

\begin{figure*}[!t]
\centering
\includegraphics[width=1.0\textwidth]{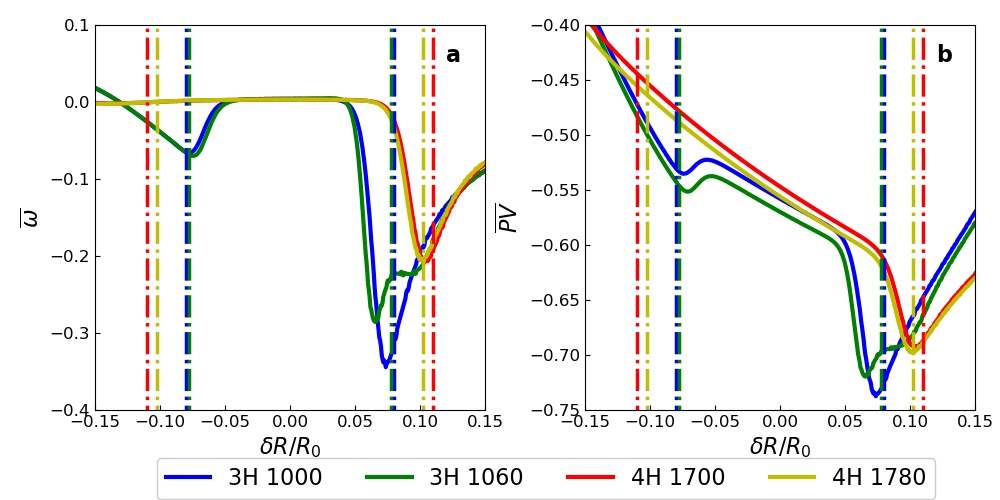}
\caption{Similar with the Figure~\ref{fig:Vorticity_PV_2H}, but for the $3H$ and $4H$ models. We chose the frames without instability (``$3H$ 1000'' and ``$4H$ 1700'') and the frames with instability (``$3H$ 1060'' and ``$4H$ 1780'').}
\label{fig:Vorticity_PV_3H_4H}
\end{figure*}

The general behaviour of the gas and dust evolution for disk models with different values of $\Delta_\textrm{DZ}$ is presented in Figure~\ref{fig:hydro_summary}. The imposed radial variation of the viscosity parameter $\alpha$ (Equation~\ref{eq:alpha}) leads to the formation of local maxima in the gas pressure between $R/R_0=0.8$ and 1.1 (panel a). As time passes, dust particles drifting inward from the outer disk regions are trapped within the gas pressure maxima, resulting in a strong radial and temporal variation of the dust-to-gas ratio (panel b and c). At the end of the simulation, the azimuthally averaged dust-to-gas ratio at the position of the dusty rings approaches unity in all three models (panel b and c).

The temporal evolution of the dust-to-gas ratio strongly depends on the radial profile of $\alpha$. In the $2H$ model, the maximum value of the dust-to-gas ratio increases above 1 after about 300 orbits and shows large temporal oscillations caused by the generation and dissipation of small dense clumps of dust within the gas ring (Figure~\ref{fig:density_2H_1} and \ref{fig:density_2H_2}). We will discuss these structures in detail in the next section. Conversely, the maximum of dust-to-gas ratio in model $3H$ remains below about 2 for the entire simulation and shows small temporal variation, while, in model $4H$, the maximum dust-to-gas ratio settles around a value of 0.7 after reaching unity
after about 500 orbits. The total mass of solids also evolves with time and it decreases from the initial value of 45~$\mrm{M_\earth}$ to about 35 $\mrm{M_\earth}$, 42 $\mrm{M_\earth}$ and 43 $\mrm{M_\earth}$ at the end of simulation in the $2H, 3H$ and $4H$ models, respectively. Note that all the dust particles in the outer disk are collected in the ring.

Figure~\ref{fig:density_2H_1} to \ref{fig:density_4H} show the temporal evolution of the gas and dust surface density in the $2H, 3H$, and $4H$ models. A common feature among models is the formation of ripples at the outer edge of the dusty rings, which, as discussed below, we attribute to the emergence of a meso-scale instability. Both the time scale for exciting the instability and their amplitude strongly vary from model to model based on the steepness of the viscosity transition. The origin of these features is intrinsically related to the feedback that dust particles induce on the gas dynamics.

If dust feedback is not accounted for, dust grains would be passively transported by the gas and would concentrate within the gas pressure bump until balanced by dust diffusion \citep{takeuchi2002radial}. Without the dust feedback, the dusty ring would be very narrow and dense, and become gravitational unstable if self-gravity is included. The maxima of the dust-to-gas ratio would easily exceed unity. Conversely, if dust feedback is considered, the gas dynamics will be significantly affected by solid particles as soon as the dust-to-gas ratio approaches unity. This leads to a substantial change in the distribution and kinematics of the gas that might hinder building up of pressure bump \citep{taki2016dust}. And the widths of dusty rings become broader~\citep{kanagawa2018impacts} including dust feedback. Our simulations indicate that dust feedback leads to an unstable rings and might results in the formation of dense clumps. The details of these instabilities are discussed in the following sections.

\subsection{$2H$ model}
Figure~\ref{fig:density_2H_1} and \ref{fig:density_2H_2} present the temporal evolution of the gas and dust surface density in the $2H$ model. Around orbit 260, the dust-to-gas ratio at the outer boundary of the dusty ring reaches unity, and five billow-shaped features emerge in the dust distribution. Around orbit 600, both the inner and outer boundaries of dusty ring became unstable and feature the formation of small vortices that grow in number as the dust trapping becomes stronger. At orbit 1740, more than 50 dust clumps with dust-to-gas ratio as high as 50 are visible along the dusty ring. The dust clumps have masses between 0.1 and $0.5\;\mrm{M_\earth}$. They are roughly elliptical in shape and have a characteristic diameter of 0.6 au, or 1/5 of the disk scale height. Our simulations resolved the dust clumps with about 10 grid cells. As time proceeds, the strong dust feedback destroys the vortices \citep{fu2014effects} and, after about 2740 orbits, the dusty ring returns to be almost featureless with a dust-to-gas ratio around 2. After that, the disk undergoes a new phase of instabilities that lead to the generation and dissipation of new small vortices. In the last part of our simulation, between orbit 3000 and 4000, the dusty ring features dense dust clumps and feather-like features that extend over $180^\circ$ along the azimuthal direction.

\subsection{$3H$ and $4H$ models}
Figure~\ref{fig:density_3H} and \ref{fig:density_4H} present the temporal evolution of the gas and dust surface density for models $3H$ and $4H$, respectively. As shown in Figure~\ref{fig:hydro_summary}, shallower variations of $\alpha$ result in pressure bumps that are both shallower and located closer to the central star. In both model, the outer edge of the dusty ring becomes unstable. However the instability grows at a lower pace and to a lower amplitude compared to the $2H$ model.

In model $3H$, billow-shapes features appear at the outer edge of the ring after about 1780 orbits, however the instability never leads to the formation of small scale vortices as those observed in Figure~\ref{fig:density_2H_1}. Furthermore, no instabilities develop at the inner edge of the dusty ring. In model $4H$, weak billow-shapes features appear around orbit 1780 and remain quasi-stable for the entire length of the simulation. At orbit 4000, the azimuthally averaged dust-to-gas ratio in the $3H$ and $4H$ model, is 0.8 and 0.6, respectively.

\subsection{Origin of Instability}
To understand the instability revealed in the simulation results described above, we studied the gas and dust velocity profiles in more detail, which is shown in Figure~\ref{fig:origin}. A circumstellar disk characterized by a monotonic decreasing profile of the gas surface density and temperature typically rotates at a slightly sub-Keplerian velocity \citep{adachi1976gas,weidenschilling1977aerodynamics}.  For the assumed initial disk parameters, the gas orbital velocity $V_\mrm{g,\Phi}$ at $R_0$ is about $1-O\left(\frac{H}{R}\right)^2$ of the Keplerian velocity $V_\mrm{K}$. As a radial maximum in the gas pressure forms as the result of the assumed viscosity profile, the gas located inside the pressure maximum is forced to rotate faster than $V_\mrm{g,\Phi}$, while the gas outside the pressure bump slows down. This variation in gas velocity forces dust particles to drift toward the pressure maximum and the dust-to-gas ratio to grow. Neglecting feedback of dust particles on the gas dynamics, and without any other process capable to halt the dust drift, solid particles will smoothly concentrate in an very narrow dusty ring without affecting the rotational velocity of the gas. However, when the dust feedback is included, as the dust density increases, solid particles would push the gas to rotate a Keplerian velocity creating inflection points in the gas velocity and vorticity profile. As the dust-to-gas ratio within the dusty ring approaches unity, the gas in the ring will be ``forced'' to rotate almost at the Keplerian velocity, while the gas just outside the ring, where the dust-to-gas ratio is much smaller, will still move at sub-Keplerian velocity. This leads to a steep velocity shear that makes the outer edge of the dusty ring unstable.
The azimuthally averaged orbital vorticity ($\omega = \left[\nabla \times \left(\mbf{V}_\mrm{g} - \mbf{V}_\mrm{K}\right)\right]_z$) and potential vorticity ($PV$, vortensity) profiles where $PV = \omega/\Sigma_\mrm{g}$ for $2H$ are shown in Figure~\ref{fig:Vorticity_PV_2H}. Same profiles for $3H$ and $4H$ models are shown in Figure~\ref{fig:Vorticity_PV_3H_4H}, respectively. The three curves in Figure~\ref{fig:Vorticity_PV_2H} show the development of instability (orbit 200, 260) before it becomes completely nonlinear (orbit 600). Ultimately, the velocity knee at the outer edge the disk smooths out, eventually leading to a relative quiescent phase around orbit 2740. Similar behaviors are captured for $3H$ and $4H$ models as shown in Figure~\ref{fig:Vorticity_PV_3H_4H}.

Figures~\ref{fig:origin} - \ref{fig:Vorticity_PV_3H_4H} indicate that the dust feedback is essential in producing the sharp features in the $PV$ profiles near the edges of the dusty rings in all cases. We speculate that such profiles will render the flow unstable to the Rossby wave instability \citep{lovelace1999rossby,li2000rossby,li2001rossby,ono2016parametric,ono2018parametric}. In fact, Figure~\ref{fig:density_2H_1} to \ref{fig:density_4H} show that the linear development of unstable modes have an azimuthal mode number between 3 and 5, similar to the predictions given in \cite{li2000rossby} for the most unstable modes. We plan to carry out more detailed linear theory analysis of this instability in a future study. For now, we will keep referring to the instability as a ``meso-scale" instability because the features generated by the instability have length-scales ranging from $\sim H$ to tens of $H$.

Density waves excited by planets~\citep{goldreich1978excitation,goldreich1979excitation,goldreich1980disk} carry angular momentum flux~\citep[AMF,][]{goodman2001planetary,rafikov2002planet,dong2011density1,dong2011density2,bae2018planet1,bae2018planet2,miranda2019gaps,miranda2019multiple} while propagating. When the density waves shock, AMF is transferred into the disk and gaps can be opened in an inviscid disk eventually. Because gap edges are also pressure bumps, similar instabilities could happen at dusty rings induced by planets including dust feedback. \cite{pierens2019vortex} showed interactions between a low-mass planet and an inviscid pebble-rich disk. They found that a RWI-liked instability happens at gap edges of planet and a lot of dusty vortices emerged in their simulations. \cite{yang2019morphological} studied planet-disk interactions including dust feedback. They showed that large-scale lopsided vortices on either gap edge of planets would be broken into numerous small dusty vortices. When the small dusty vortices dissipate, the large-scale vortices reappear again. We believe that both what they found are consistent with the instability in this paper.

\section{Synthetic ALMA images} \label{sec:images_results}
\begin{figure*}[!t]
\centering
\includegraphics[width=1.0\textwidth]{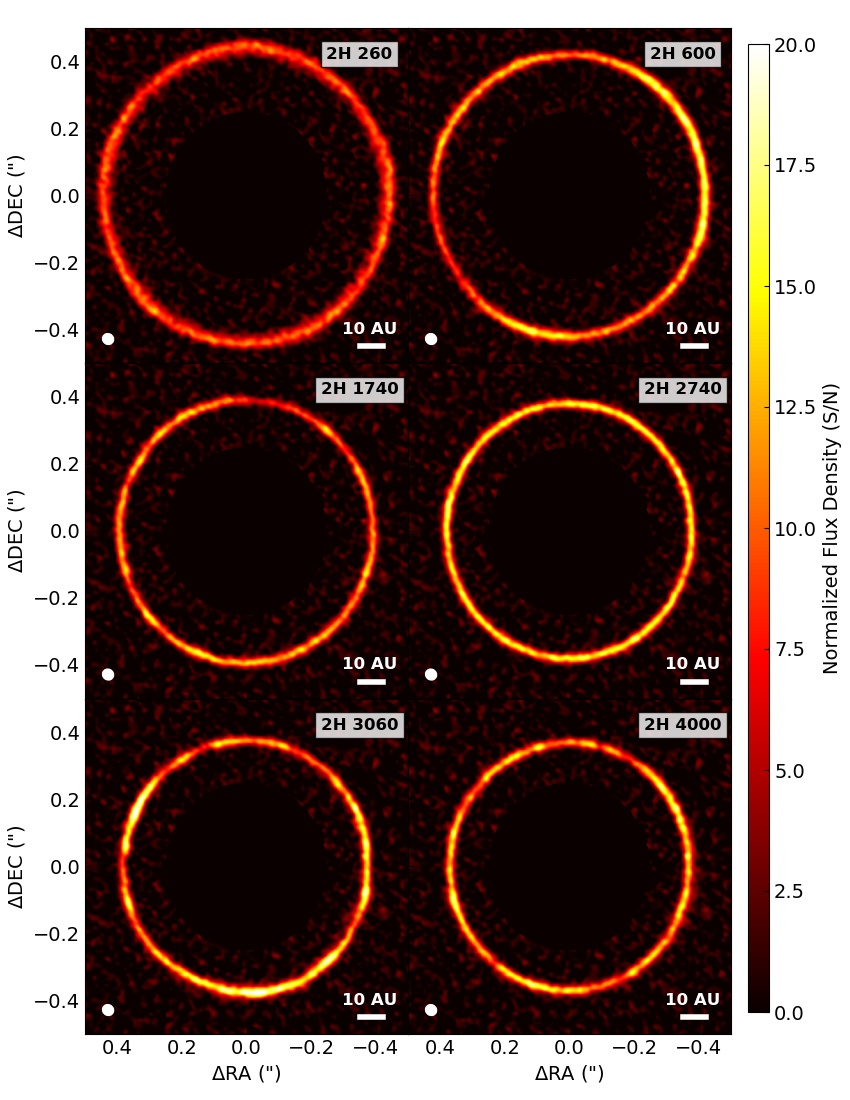}
\caption{The synthetic ALMA images of specific orbits at 1.3 mm wavelength for $2H$ model. The intensity is normalized by signal-to-noise ratios (S/N). The x and y axes present the relative location from the central star along right ascension and declination. The synthetic beam is marked as a white ellipse. The FWHMs of synthetic beams are $0.039\arcsec$ and $0.036\arcsec$, i.e. the spatial resolution is $5.5\times5.0\;\mrm{au}$. The RMS of noise is about $6.1 \;\mrm{\mu Jy\; beam^{-1}}$.}
\label{fig:image_2H}
\end{figure*}

\begin{figure*}[!t]
\centering
\includegraphics[width=1.0\textwidth]{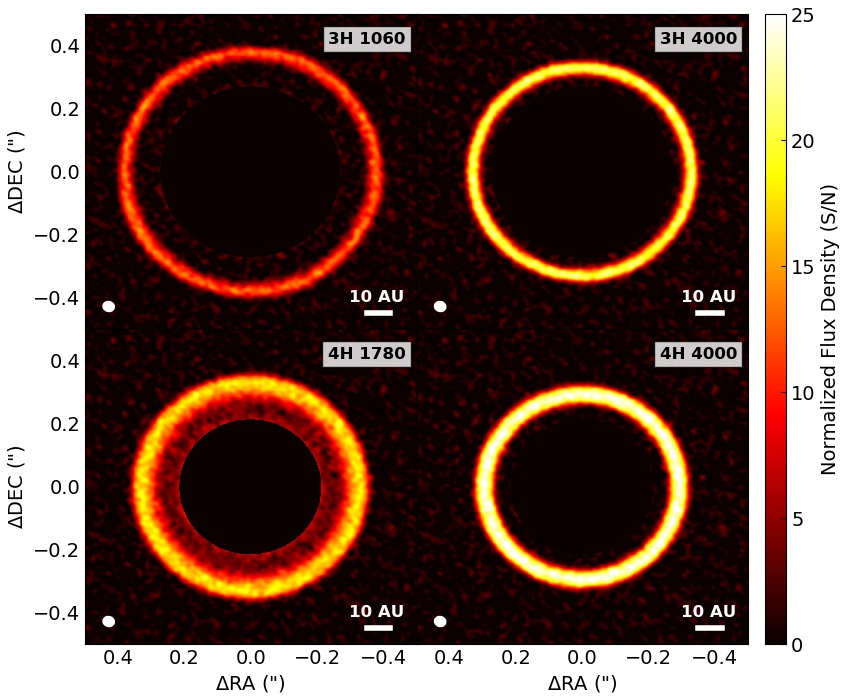}
\caption{The synthetic ALMA images of specific orbits at 1.3 mm wavelength for $3H$ and $4H$ models. The beam size and the noise level are same as the $2H$ model in Figure~\ref{fig:image_2H}. Because the instability is quite weak in $3H$ and $4H$, we just chose the frames with instability occurring and last frames of simulations. We can see that the rings in $3H$ and $4H$ are quite smooth compared with rings in $2H$.}
\label{fig:image_3H_4H}
\end{figure*}

\begin{figure*}[!t]
\centering
\includegraphics[width=1.0\textwidth]{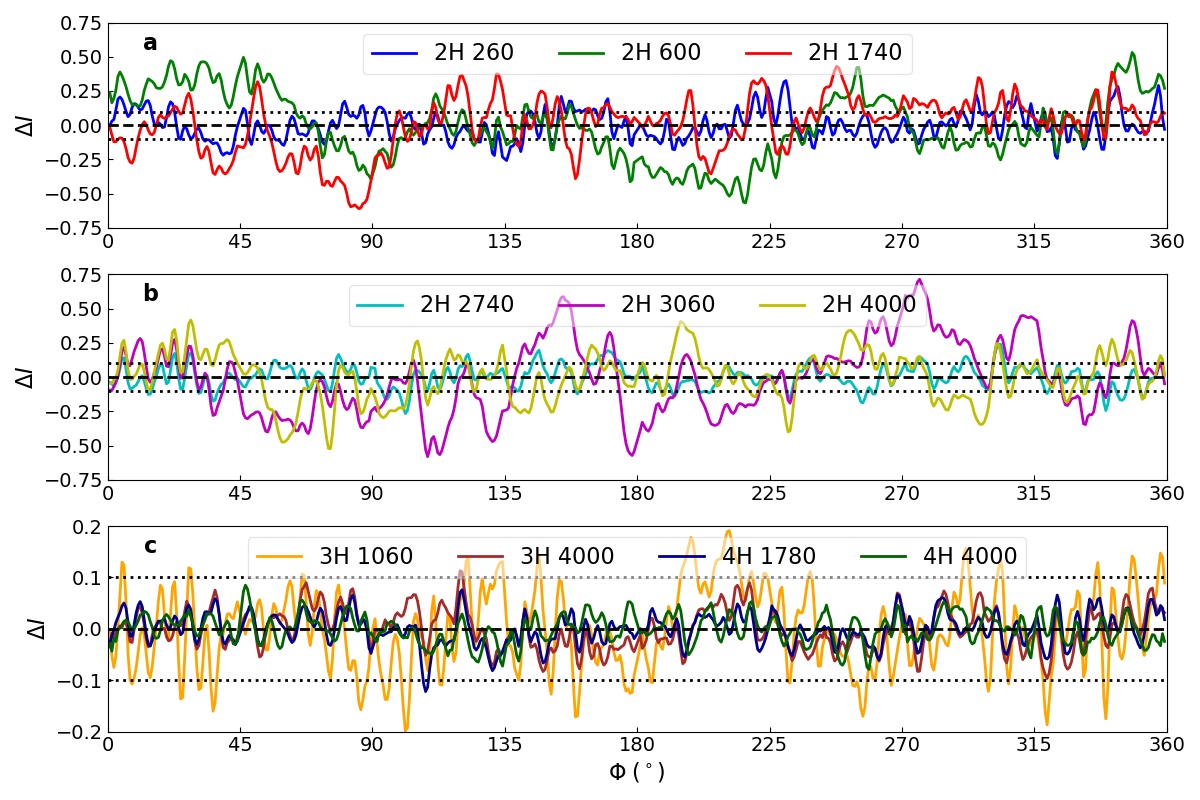}
\caption{Normalized intensity variation along the dusty rings for $2H$, $3H$ and $4H$ models. The black dashed lines mark the variations equaled to zero. The normalized intensity variation is defined as $\Delta I = (I - \overline{I})/\overline{I}$, where $\overline{I}$ is the azimuthally averaged intensity along dusty rings. The definition of the azimuthal angle $\Phi$ is started from right with anti-clockwise direction (from the west to the north), same as Figure~\ref{fig:density_2H_1} - \ref{fig:density_4H}, Figure~\ref{fig:image_2H} and \ref{fig:image_3H_4H}. The radial width of azimuthal cut is about 1 radial gaussian FWHM width for each dusty ring. For convenience, we calculate the variations of rings from intensity images directly, instead of calculating visibility on uv-plane. The amplitude of fluctuations caused by noise is an order of $1/\left(S/N\right)\sim 0.1$, as the black dotted lines show in each panel.}

\label{fig:delta_intensity}
\end{figure*}

\begin{figure*}[!t]
\centering
\includegraphics[width=1.0\textwidth]{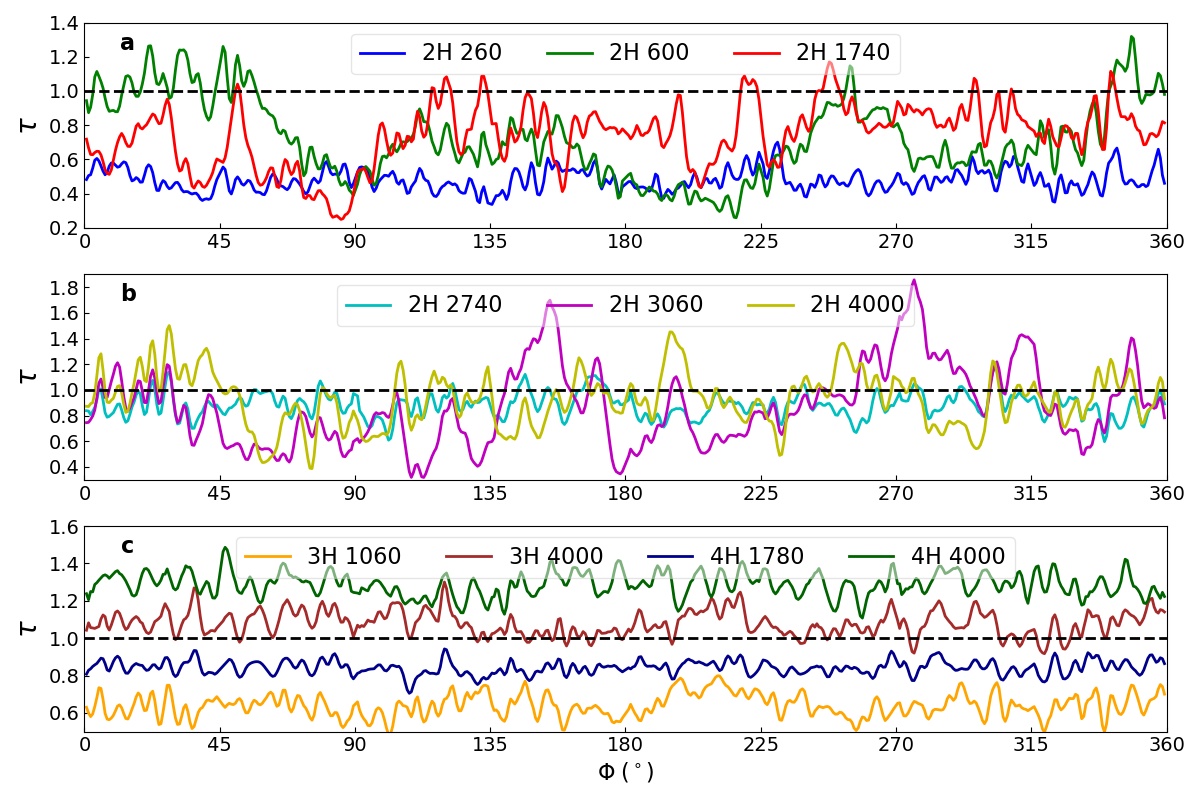}
\caption{The optical depth $\tau$ of specific orbits for $2H$, $3H$ and $4H$ models. The black dashed lines mark the optical depth equaled to unity. The small fluctuation of optical depths is caused by noise. The optical depth of dusty rings for $2H$ present a large oscillation by strong instabilities. The optical depth of dusty rings for $3H$ are about 0.6 and 1.1 at 1060 and 4000 orbits, respectively. The optical depth of dusty rings for $4H$ are about 0.8 and 1.2 at 1780 and 4000 orbits, respectively.}
\label{fig:delta_tau}
\end{figure*}

\begin{figure}[!t]
\centering
\includegraphics[width=0.5\textwidth]{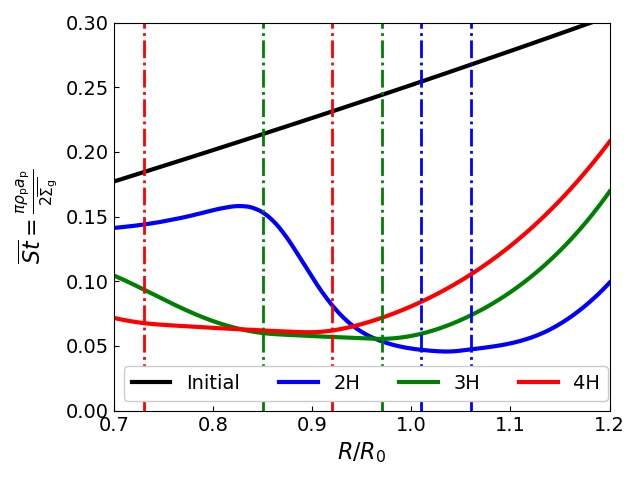}
\caption{Azimuthally averaged Stokes number at 4000 orbits for $2H$, $3H$ and $4H$ models. The black line is the Stokes number of the initial frame and the dashed-dotted lines marked the boundaries of the dusty rings for each model. The Stokes numbers in the dusty rings are about 0.045, 0.055 and 0.060 for $2H$, $3H$ and $4H$, respectively.}
\label{fig:stokes_number}
\end{figure}

\begin{figure}[!t]
\centering
\includegraphics[width=0.50\textwidth]{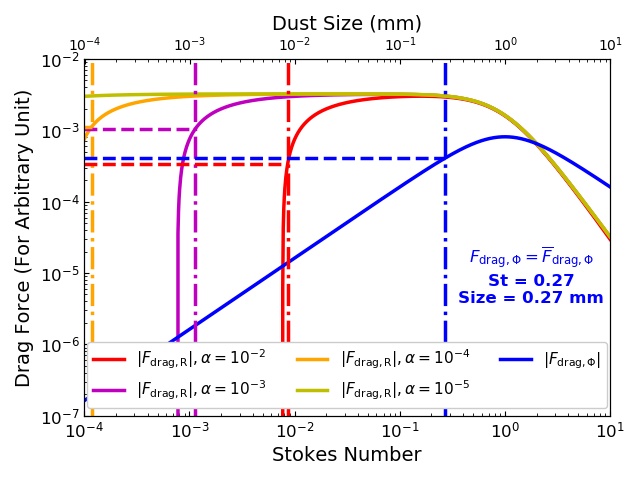}
\caption{The radial and azimuthal components of drag force in the equation~\ref{eq:drag} varied with the Stokes number and dust size for different $\alpha$. The x axes are the Stokes number and dust size. The y axis is the strength of the drag force in arbitrary unit. The dashed lines and dashed-dotted lines mark the Stokes numbers and dust size when the drag force equal the mean drag force.}
\label{fig:drag_force}
\end{figure}

In this section, we investigate whether ALMA high angular resolution observations of nearby circumstellar disks could detect some of the small scale structures resulting from the instability.

\subsection{Intensity of the continuum emission}

Figures~\ref{fig:image_2H} and \ref{fig:image_3H_4H} show synthetic images of different snapshots for $2H$, $3H$ and $4H$ models in the 1.3 mm dust continuum emission calculated as discussed in Section~\ref{sec:setup}. The spatial resolution of the observations is about 5 au and is comparable to the radial extent of the ring. The RMS noise is $6.1 \;\mrm{\mu Jy\;beam^{-1}}$ and the emission is detected with a peak signal-to-noise ratio (S/N) of $20\sim25$. At this sensitivity and resolution, none of the billow-shaped features that characterize the edges of the dusty ring in $2H$ model are visible. Also, none of the dense dust clumps is singularly resolved. However, the synthetic images show azimuthal variations in the intensity profile related to the instability in the $2H$ model. Models $3H$ and $4H$ are characterized by a lower level of instability which lead to more azimuthally symmetric dusty rings. Only at orbit 1060 in the $3H$ model, the azimuthal variations in the intensity marginally trace the location of the billow-shapes structures present at the outer edge of the ring. At all the other orbits in both models, the intensity variations are consistent with the noise of the observations.

More detailed analysis of the $2H$ model reveals the following features: At orbit 260, the azimuthal intensity profile shown in Figure~\ref{fig:delta_intensity} oscillates between $\pm20\%$ of the mean intensity in accordance with the position of the billow-shaped features shown in \ref{fig:density_2H_1}. Essentially, at the azimuthal positions where these features occur, the dusty ring is more extended radially and its radially integrated flux is larger. We argue that a power spectrum analysis of the azimuthal intensity profile of known dusty rings might reveal the presence of the instability. At orbit 600, the azimuthal intensity profile shows variations as large as 50\% which are caused by the non-uniform distribution in azimuth of the dense dust clumps. Indeed, at this orbit, the dust clumps are organized in two clusters centered around $\Phi=30^\circ$ and $\Phi=250^\circ$, containing about 10 dust clumps each. At orbit 1740, the dust surface density is characterized by a large number of clumps, which are more uniformly distributed along the ring compared to the previous case, but nevertheless lead to intensity variations of about 30\% of the mean intensity. A similar behaviour of the intensity profile is observed at orbits 2740, 3060, and 4000, in which amplitude variations correlate to the amount of instabilities in the disk and can lead to a contrast of intensity along the ring ($I_\mrm{max}(\Phi)/I_\mrm{min}(\Phi) = (1+\Delta I_\mrm{max}(\Phi))/(1+\Delta I_\mrm{min}(\Phi)$) up to 4, as for orbit 3060.

\subsection{Optical Depth}
We calculate the fraction of dust mass trapped inside regions that are optically thick in the dust continuum emission at 1.3 mm. Following the usual procedure adopted to derive the optical depth of the continuum emission from ALMA observations, we calculate the optical depth $\tau$ of our synthetic intensity maps from the relation:
\begin{equation}
\label{eq:tau}
    I_\nu = B_\nu\left(T_\mrm{d}\right)\left(1 -e^{-\tau_\nu}\right),
\end{equation}
where $T_\mrm{d} \simeq 12$~K is the dust temperature inside the dusty ring, and $B_\nu$ is the Planck function. A clear advantage of working with simulated data is that we know the true value of the dust temperature, while in the case of real observations, the dust temperature is generally estimated based on radiative transfer models or observations of optically thick molecular lines. Note that the optical depth calculated using the previous equation is an average across the synthesized beam. In the presence of structures smaller than the beam, $\tau$ would therefore differ from the ``true'' optical depth of the emission as calculated by multiplying the dust surface density map by the dust opacity. The comparison between the optical depth calculated from the synthetic images and that derived from the surface density map, provides a measurement of the beam dilution.

Figure~\ref{fig:delta_tau} shows the optical depth for the $2H, 3H$, and $4H$ models calculated using Equation~\ref{eq:tau}. In the orbit 260 frame of model $2H$, the optical depth of the dusty ring varies between 0.4 and 0.7. As dense dust clumps form as a result of the instability, $\tau$ increases to a maximum of 1.3 at orbit 600. This value should be compared with the optical depth at the same position calculated from the surface density, which, in this specific case is 20. This implies that the limited angular resolution of the observations leads to a dilution of intensity of about a factor of 15, or that the structures that emit most of the emission at that specific location in the disk have a size about 15 times smaller than the angular resolution of the observations.

The $3H$ and $4H$ models have lower dust surface densities than the $2H$ model. However, because the dust is more uniformly distributed, the optical depth inferred form the simulated observations is larger than in the $2H$ model, and varies between $0.6\sim1.1$ and $0.8\sim1.2$, respectively. The maximum optical depth of the $3H$ model as calculated from the dust surface density map is about 2.2, implying that beam dilution plays a much smaller role than that in the $2H$ models. This is consistent with the lack of dense clumps in this model. Even smaller is the effect of beam dilution in the $4H$ model, where the maximum ``true'' dust optical depth peaks at 1.8.

A direct consequence of the presence of optically thick and spatially unresolved dust clumps is that the dust mass estimated from the observed intensity will underestimate the true amount of dust trapped in the dusty ring. For example, the synthetic observations of model $2H$ provide a total dust mass between 11 and 18 $\mrm{M_\earth}$, while the true dust mass trapped within the ring is between 20 and 35 $\mrm{M_\earth}$ in $2H$ model. In the case of the $3H$ model, the dust mass inferred from the observations is between 18 and 27 $\mrm{M_\earth}$, to be compared to a true dust mass between 30 and 42 $\mrm{M_\earth}$. Finally, for model $4H$, the mass estimated from the observations is between 35 and 37 $\mrm{M_\earth}$, to be compared with a true mass between 37 and 42 $\mrm{M_\earth}$.

These results stress the importance of imaging dusty rings at the highest angular resolution possible to mitigate the effect of beam dilution, and at wavelengths longer than 1 mm to reduce the limiting effect of the dust optical depth.

\section{Discussion} \label{sec:discussion}

\subsection{Dependence on Model Parameters}\label{subsec:parameters}
The results discussed above rely on several assumptions concerning the origin of the radial dust trap,
the initial dust and gas densities, the dust grain size and opacity, the equation of state, the lack of self-gravity, and the fact that we used 2D simulations whereas circumstellar disks are 3D objects. Here we discuss how some these assumptions might affect our results and present ideas for future follow-up studies.

\subsubsection{Dust Size and Stokes Number}\label{subsubsec:stokes}
Dust feedback is a necessary condition for the instability found in this study. The dust feedback depends on the coupling between gas and dust, which, in turn, is controlled by the dust Stokes number. Our simulations assume a single grain size $a_\mrm{p}=0.25$ mm, corresponding to a Stokes number of about 0.25 at $R_0$. As a result of the radial evolution of the gas density, the Stokes number varies with time and radius and, at the end of the simulations (at orbit 4000), assumes values between 0.05 and 0.15 across the dusty ring (Figure~\ref{fig:stokes_number}).

It is important to understand how different sizes of the particles contribute to the dust feedback. To do that, we adopt the approach of \cite{takeuchi2002radial} and express the difference between the azimuthal velocity of gas and dust as
\begin{equation}
\label{eq:dvphid}
V_{\mrm{d,\Phi}}-V_{\mrm{g,\Phi}}= -\frac{1}{2}V_{\mrm{d,R}}St,
\end{equation}
where $V_\mrm{d,R}$ is the radial velocity of the dust
\begin{equation}
\label{eq:vrd}
    V_{\mrm{d,R}} =-\frac{ \eta V_\mrm{K}}{St + St^{-1}},
\end{equation}
and
\begin{equation}
\eta=-\left( \frac{H}{R} \right)^2 \frac{\partial \ln P}{\partial \ln R}
\end{equation}
describes how much slower the gas rotates compared to the Keplerian velocity due to the radial pressure gradient. Combining these two equations with Equation~\ref{eq:drag}, we find
\begin{equation}
F_{\mrm{drag,\Phi}} = \frac{\eta \Omega_\mrm{K} V_\mrm{K}}{2(St+St^{-1})}
\end{equation}
Along the radial direction, the velocity difference between gas and dust is
\begin{equation}
\label{eq:dvrd}
    V_\mrm{d,R}-V_\mrm{g,R} = -\frac{\eta V_\mrm{K}}{St+St^{-1}} + \frac{\alpha c_\mrm{s} H}{R},
\end{equation}
which combined to Equation~\ref{eq:drag} gives
\begin{equation}
    F_\mrm{drag,R} = \frac{\eta V_\mrm{K} \Omega_\mrm{K}}{St(St+St^{-1})}- \frac{\alpha c_\mrm{s}^2}{R St}.
\end{equation}

If the dust population is composed of grains with different sizes, the average of the dust drag force per unit of mass is
\begin{equation}
    \overline{F}_\mrm{drag,R} = \frac{\int_{a_{min}}^{a_{max}} F_\mrm{drag,R} n(a) a^{3} da}{\int_{a_{min}}^{a_{max}}n(a) a^{3} da},
\end{equation}
\begin{equation}
    \overline{F}_\mrm{drag,\Phi} = \frac{\int_{a_{min}}^{a_{max}} F_\mrm{drag,\Phi} n(a) a^{3} da}{\int_{a_{min}}^{a_{max}}n(a) a^{3} da},
\end{equation}
where $n(a) \propto a^{-3.5}$ between $a_{min} =0.1$ $\mu$m ($St = 10^{-4}$ at $R_0$) and $a_{max} = 1$ cm ($St = 10$ at $R_0$) is a typical grain size distribution for circumstellar disks. We numerically integrate the latter equation and calculate the characteristic particles size by solving the relations $F_\mrm{drag,R}=\overline{F}_\mrm{drag,R}$ and $F_\mrm{drag,\Phi}=\overline{F}_\mrm{drag,\Phi}$. As Figure~\ref{fig:drag_force} shows, the drag force along radial direction is dominated by the small Stokes number (small dust size). And the drag force along the azimuthal direction is dominated by the dust with $St\simeq 0.27$. For the assumed initial profiles of $\alpha$ and $c_\mrm{s}$, we find that the characteristic Stokes number of 0.27 along the $\Phi$ direction corresponding to a grain size of 0.27 mm, which is very close to the one adopted in our simulations. Initializing the simulations with a larger grain size (larger than 0.27), would have reduced the dust feedback and might have led to weaker instabilities.

Our findings are also consistent with some previous studies. The ``ViscBroad'' model in Figure 4 of M17 is similar with our $2H$ model. However, the dusty ring in ``ViscBroad'' just presents weak asymmetries and does not show a similar instability compared with $2H$ model. Because M17 used the dust particle with initial Stokes number equaled to unity, this is consistent with our analysis above. \cite{fu2014effects} found that dust will be concentrated effectively by Rossby Wave vortices and dust feedback will destroy vortices when the dust density is comparable and higher than the gas density. They used dust with Stokes number from $4\times 10^{-4}$ to 0.16 in their simulations. They found that the dust with large Stokes number has a large back-reaction on the vortices. This is consistent with the fact that the small vortices emerge and dissipate in hundred orbits of our simulations.

In the paper, we just use one size dust component in our simulations. The simulations on full dust size distribution will be quite costly computationally. And we do not consider dust coagulation and fragmentation in the paper, which could play an important role in the dust evolution in the protoplanetary disk~\citep{drazkowska2019including}.

\subsubsection{2D approximation, Dust to Gas Ratio and Viscosity Profile}\label{subsubsec:ratio_viscosity}

We have used 2D approximations in our simulations. The initial instability wavelength is significantly longer than $H$ though the transition width is on the order of $H$. Some of the non-linear features generated in the $2H$ model can have sizes less than $H$. We used about 26 cells to capture these small vortices. On the other hand, features with size $< H$ should be taken as indicative only because
2D approximations will not be fully consistent with such features.

In this study, we used three models with different viscosity transition widths to generate pressure bumps to trap dust. We find that the instability gets triggered when the dust-to-gas ratio approaches unity. We use a relatively high dust-to-gas ratio at the beginning ($= 0.05$), however, it does not affect the main conclusions. In our three models, the dust-to-gas ratio in the dusty rings all exceed or approach unity. We infer the critical dust-to-gas ratio is about 0.6 based on the weak instability at the outer edge of dusty rings in $4H$. In actual disks, the dust is likely to settle towards the mid-plane, producing higher dust to gas ratio. The processes that are studied here will likely to be applicable to the mid-plane region, though detailed 3D studies have to be carried out to test this scenario.

\subsubsection{Dust Opacity, Optical Depth and Dust Scattering}\label{subsubsec:opacity}
The dust opacity at (sub-)millimeter wavelengths in protoplanetary disk is commonly assumed as $\kappa \propto \nu^{\beta}$, where $0< \beta \lesssim 1$ \citep{beckwith1990survey} and we adopt $\beta = 0.4$ in this paper. We calculate the optical depth of the rings in our synthetic images and find that the optical depth is between 0.4 and 1.4 at 1.3 mm wavelength with $\sim 0.04\arcsec$ resolution for our models. Note that the width of dusty rings in $2H$ ($\sim 3$ au) is smaller than the beam size ($\sim 5$ au). The local optical depth of the dust clumps in $2H$ model are optically thick without the dilution of observational beam. The dusty rings will be optically thick at higher frequency and optically thin at lower frequency, respectively. When dusty rings become optically thick, the intensity variation (see Figure~\ref{fig:delta_intensity}) of dusty rings will become small and the dust clumps will be hard to distinguish.

The opacity of dust can be affected significantly by temperature, chemical composition, grain size, structure and topology of dust particle~\citep{semenov2003rosseland}. The dust component in protoplanetary disks is different with the interstellar grains~\citep{draine2006submillimeter}. The uncertainties of dust opacity also comes from the opacity models and the numerical computing~\citep{birnstiel2018disk}. There could be an order of variation on the dust opacity for different models~\citep{demyk2017low1,demyk2017low2}. For optically thin medium, the scattering can be ignored. In the situation with intermediate and high optical depth, the scattering opacity at (sub-)millimeter wavelength will create polarization emission of disks~\citep{kataoka2015millimeter} and extinct part of CO emission~\citep{isella2018disk}. \cite{liu2019anomalously} showed that the self-scattering of optically thick dust disk can produce low (sub-)millimeter spectral indices. \cite{zhu2019one} proposed that optically thick dust with scattering can lead to underestimating dust masses. The effects of scattering on interpreting our simulation images will be explored in the future.

\subsection{Self Gravity, Gravitational Instability and Streaming Instability}\label{subsubsec:GI}
When the thermal pressure and the rotation of disk can not support the self-gravity of disk, the gravitational instability (GI) will happen. For gas disk, this can be characterised by Toomre Q value~\citep{toomre1964gravitational,goldreich1965gravitational}:
\begin{equation}
Q \equiv \frac{\kappa c_\text{s}}{\pi G \Sigma}
\label{eq:toomre_Q}
\end{equation}
where $\kappa$ is the radially epicyclic frequency ($\kappa^2\equiv \frac{1}{R^3} \frac{d}{dR}\left(R^4 \Omega^2\right)$), and $G$ is gravitational constant. For pure dust disk (such as Saturn rings), the velocity dispersion $\sigma$ between dust particle replaces the sound speed $c_\text{s}$ in Equation~\ref{eq:toomre_Q}~\citep{armitage2007lecture}. When $Q$ is smaller than 1, GI will happen when the self-gravity is included.

\begin{figure*}[!t]
\centering
\includegraphics[width=1.0\textwidth]{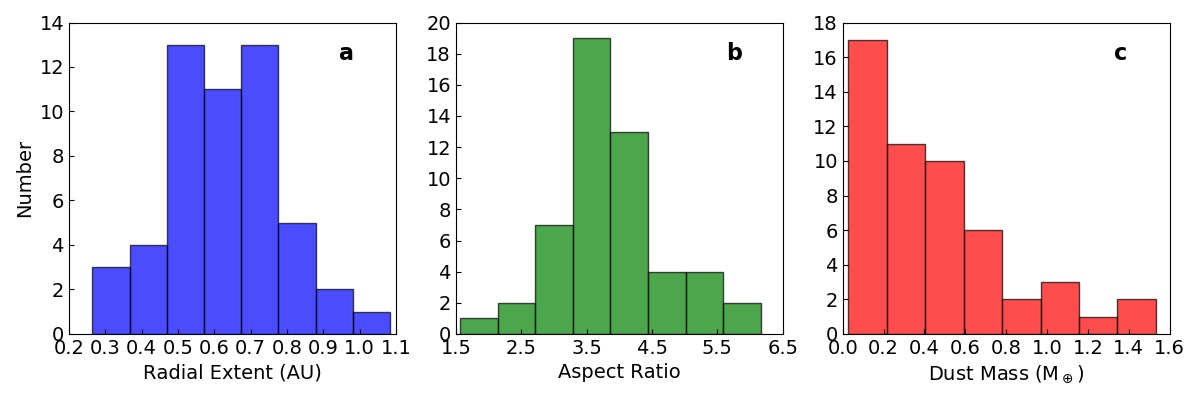}
\caption{The histograms of properties of dust clumps in orbit 1740 for $2H$ shown in Figure~\ref{fig:density_2H_1}. Panel a: Radial extents of dust clumps. Panel b: Aspect ratios between $\Phi$ extent and $R$ extent of dust clumps. Panel c: Masses of dust clumps. We assume the dust clumps are with ellipse shape, then measure the peak density and densities at radial and azimuthal edges so as to estimate the masses in dust clumps. There are more than 50 clumps in the dusty ring in orbit 1740 for $2H$. The total masses of dust clumps are about $23\;\mrm{M_\oplus}$, compared with $34\;\mrm{M_\oplus}$ for the whole dusty ring. The medians of radial extents, aspect ratios and dust masses are 0.6 au, 3.6 and $0.4\;\mrm{M_\oplus}$.
}
\label{fig:hist_clumps}
\end{figure*}

Dust is trapped in the small vortices created by the instability where dust-to-gas ratio is high ($\Sigma_\text{d}/\Sigma_\text{g} \gtrsim 10$), such as ``$2H$ 1740'', ``$2H$ 3060'' and ``$2H$ 4000'' (see panel $c$ of Figure~\ref{fig:hydro_summary}).  Because dust are marginally coupled with gas ($St \lesssim 0.1$ in dusty rings, see Figure~\ref{fig:stokes_number}), we can still use the sound speed $c_\text{s}$ to calculate $Q$ as an approximation. Figure~\ref{fig:hist_clumps} describes the overall properties of the dust clumps developed at late time in the $2H$ model. We can see that the median values of radial extent, aspect ratio and masses of dust clumps are 0.6 au, 3.6 and 0.4 $\mrm{M_\earth}$, respectively. The Toomre $Q$ of the dust clumps is then about 5, so the dust clumps are marginally gravitational unstable. Since we did not include self-gravity in our simulations, it might be interesting to explore
the effects of GI in determining the properties of these dust clumps. Planetesimals might be formed in the dust clumps by the aid of GI.

The former studies \citep{goodman2000secular,youdin2005streaming,youdin2007protoplanetary,johansen2007protoplanetary} showed that dust can be concentrated under the dust-gas mutual aerodynamic drag, leading the local, linear Streaming Instability (SI). SI can generate high dust-density enhancement without self-gravity. The instability found in this paper is on much larger scale (at least initially), so we believe that this instability is different from the traditional SI. The nonlinear outcome seen in the $2H$ model, however, could interplay with the SI. In the simulations presented here, we do not have enough radial resolution per $H$ to capture the SI, so presumably the traditional SI is not excited here. The vertical motion of dust is also important for SI, so interplay between this instability and SI still need to be carried out in 3D simulations in the future.

\subsection{Implication of ALMA Observations}\label{subsec:implication}
Dusty rings are the most common features in the dust continuum of protoplanetary disks~\citep{andrews2018disk,long2018gaps}. Here we discuss some protoplanetary disks with significant ring-features observed by ALMA, in the context of the instability presented in this paper.

\subsubsection{HD~163296}\label{subsubsec:HD163296}
HD~163296 is a Herbig Ae star located at the distance of 101 pc~\citep{brown2018gaia}. It has extended dust and gas disks around it~\citep{isella2016ringed,isella2018disk}. HD~163296 disk is the third circumstellar disk which is revealed by ALMA with multiple rings in dust continuum observation (after HL~Tau \citep{almapartnetship2015} and TW~Hya \citep{andrews2016ringed}). \cite{isella2007millimeter} used SMA and VLA observations to show asymmetric dust continuum and CO isotopomers depletion in HD~163296 disk. \cite{isella2016ringed} showed that there are three deleted dust gaps and reduced CO gas density in the dust gaps compared with the surroundings regions. The dust gaps and reduced CO density can be explained by the dynamical clearing by three embedded Saturn-like planets. \cite{liu2018new} used three half-Jovian planets located at the dust gaps to give a reasonable fit to the dust continuum and CO gas emission. They also claimed that HD~163296 presents a low-viscosity ($\alpha < 10^{-4}$) in the inner part and large-viscosity ($\alpha \sim 7.5\times 10^{-3}$) in the outer part of disk. The viscosity transition is consistent with the traditional dead-zone model caused by Ohmic dissipation~\citep{gammie1996layered}.

\cite{isella2018disk} presented that there are three concentric dusty rings in the HD~163296 with the Gaussian radial widths 8.7, 6.6 and 5.8 au. They also found some asymmetric features in the dust emission of HD~163296 and a significant Arc-like feature in the inner side of the dusty ring at 67 au (Figure 1 of their paper). There is about $\pm 15\%$ amplitude intensity variation relative to the mean ring intensity in the dusty rings of HD~163296. \cite{zhang2018disk} run 0.15 $\mrm{M_J}$ planet-disk interaction model with 2D hydrodynamic simulation and radiative transfer post processing. They showed that the Arc-like feature can be explained by dust trapping in the Lagrangian points along the orbit of planet. They also argued that the intensity variation of the dusty rings of HD~163296 is caused by the perturbation of planet on the gas kinematics in the disk.

From our simulations, we can see the bright clumps and crescent features along the dusty rings in our dust emission images of the $2H$ model (see ``$2H$ 1740'', ``$2H$ 3060'' and ``$2H$ 4000'' in Figure~\ref{fig:image_2H}). The dust clumps and crescents are along the dusty rings in $2H$. They differ from the Arc-like feature of HD~163296, which is located outside of the dusty ring. The large intensity variations for $2H$ model might not be matched well with $\pm 15\%$ intensity variation in HD~163296. Even though the purpose of this paper is not to give a fit on HD~163296, we infer that the asymmetric substructure of HD~163296 is probably not caused by the non-linear outcome of this instability.

\subsubsection{MWC~758}\label{subsubsec:MWC758}
MWC~758 is a young \citep[3.5 millions years,][]{meeus2012observations} Herbig Ae star with circumstellar disk, located at 160 pc away~\citep{brown2018gaia}. \cite{marino2015compact} showed that there are two bright bumps at sub-millimeter continuum. \cite{boehler2018complex} showed that MWC~758 presents complex morphology with ALMA Band 7 dust continuum and $^{12}$CO/$^{13}$CO gas line emission with angular resolution $0.1\arcsec \sim 0.2\arcsec$. \cite{dong2018eccentric} showed that there are three dusty rings, two bright bumps, and one eccentric cavity based on ALMA observation at 0.87 mm with 40 mas resolution. The contrast of the south and north bright clumps is 4.4 and 10, compared with the averaged intensity at their radii.

Two bright bumps in MWC~758 are located at different dusty rings. The contrast magnitude of the south bump is similar with the contrast from our simulations ($\sim 4$ in ``$2H$ 3060'', See Section~\ref{subsec:radiative_images}). There are also some asymmetries along the inner ring (See Figure 2 and 3 in \cite{dong2018eccentric}). The north bumps with high contrast ($\sim 10$) might come from dust trapping in a larger scale vortex generated by RWI. \cite{baruteau2019dust} showed that the bright bumps in MWC~758 are the large scale vortices of RWI triggered by two massive planets in gaps. They also claimed that the inner vortex decays faster than the outer one, then it also can explain the formation of the eccentric inner ring. In our simulations, the dusty ring is also distorted in ``$2H$ 3060''. The instability and its nonlinear outcome might provide some explanation for the complex features seen in MWC~758.

\subsubsection{HL~Tau}\label{subsubsec:HL_Tau}
HL~Tau is a young T~Tauri Star located at about 140 pc from Earth~\citep{loinard2007vlba}. \cite{almapartnetship2015} used ALMA observation to present multiple bright and dark rings in HL~Tau. Their observation also showed the low spectral index in the rings, which implies that the dust might have evolved compared with the interstellar medium grain. \cite{jin2016modeling} used three Saturn mass planets to model the rings structures in the dust continuum of HL~Tau. They also calculated the optical depth at 1.0 mm of the dusty rings and they found that the bright rings (such as B1 and B2) are optically thick, and the dark rings are optically thin. \cite{carrasco2016vla} used the VLA data on HL~Tau, and they found that the brights rings of HL~Tau at the 7.0 mm are optically thin. There are some clump candidates on the B1 ring of HL~Tau (Figure 2 in their paper).

As discussed in Section~\ref{subsubsec:opacity}, the dusty rings will be smooth in the optically thick wavelength. The bright rings in HL~Tau at the wavelengths observed by ALMA are optically thick, so they are smooth and ordered. It will be difficult to assess whether there are dense clumps formed inside these rings. Observations at multiple wavelengths are essential to confirm this instability.

\section{Summary}\label{sec:summary}
We have studied a meso-scale instability caused by dust-gas interaction at the edges of dusty rings by carrying out high resolution 2D two fluid global hydrodynamic simulation, combined with the radiative transfer calculation and ALMA synthetic image processing. Here are our main conclusions:

\begin{enumerate}
    \item If the conditions are favorable, there is an instability at the edges of dusty rings produced by pressure bumps. Dust feedback alters the azimuthal gas velocity profile, producing a sharpened potential vorticity profile at the edges of dusty rings.
    \item The instability can develop when the local dust-to-gas ratio approaches unity. A ring that is being continuously ``fed" from outer disk can eventually reach such a condition, at least transiently.
    \item The non-linear outcome of such an instability can lead to many small vortices in dusty rings. These small vortices can contain large amount of dust, reaching at least $\sim 10\%$ of Earth mass within each clump.
    \item These dust clumps are highly optically thick and they might not be able to resolved by current observations at (sub-)millimeters. However, these dust clumps still can lead to both non-axisymmetric structures and intensity variations in dusty rings. This instability can be an origin of some asymmetries observed in protoplanetary disks.

\end{enumerate}

Future studies in 3D, including interaction between this meso-scale instability with other instabilities such as SI and GI, will be quite interesting. They will help to address whether dusty rings are suitable for planetesimals formation and eventually planet formation.

\acknowledgments
We thank Richard Lovelace, Douglas Lin, Andrew Youdin, Xuening Bai, Ruobing Dong, Shangfei Liu, Cong Yu for helpful discussions. This work is supported by the National Natural Science Foundation of China (grant Nos. 11773081, 11661161013, 11633009 and 11873097), the CAS Interdisciplinary Innovation Team, the B-type Strategic Priority Program of the Chinese Academy of Sciences, Grant No. XDB41000000 and the Foundation of Minor Planets of Purple Mountain Observatory. We also acknowledge the support by a LANL/CSES project. This work was partially performed at the Aspen Center for Physics, which is supported by National Science Foundation grant PHY-1607611.

\software{LA-COMPASS \citep{li2005potential,li2008type}, RADMC-3D v0.41 \citep{dullemond2012radmc}, CASA ALMA pipeline 5.0.0 \citep{mcmullin2007casa}, Astropy \citep{robitaille2013astropy}, Numpy \citep{walt2011numpy}, Scipy \citep{jones2016scipy}}

\bibliographystyle{aasjournal}
\bibliography{references}

\begin{thebibliography}{}
\expandafter\ifx\csname natexlab\endcsname\relax\def\natexlab#1{#1}\fi
\providecommand{\url}[1]{\href{#1}{#1}}
\providecommand{\dodoi}[1]{doi:~\href{http://doi.org/#1}{\nolinkurl{#1}}}
\providecommand{\doeprint}[1]{\href{http://ascl.net/#1}{\nolinkurl{http://ascl.net/#1}}}
\providecommand{\doarXiv}[1]{\href{https://arxiv.org/abs/#1}{\nolinkurl{https://arxiv.org/abs/#1}}}

\bibitem[{Adachi {et~al.}(1976)Adachi, Hayashi, \& Nakazawa}]{adachi1976gas}
Adachi, I., Hayashi, C., \& Nakazawa, K. 1976, Progress of Theoretical Physics,
  56, 1756

\bibitem[{Akiyama {et~al.}(2016)Akiyama, Hashimoto, baobabu Liu, Bonnefoy,
  Dong, Hasegawa, Henning, Sitko, Janson, Feldt, {et~al.}}]{akiyama2016spiral}
Akiyama, E., Hashimoto, J., baobabu Liu, H., {et~al.} 2016, The Astronomical
  Journal, 152, 222

\bibitem[{Alexander {et~al.}(2006{\natexlab{a}})Alexander, Clarke, \&
  Pringle}]{alexander2006photoevaporationI}
Alexander, R.~D., Clarke, C., \& Pringle, J. 2006{\natexlab{a}}, Monthly
  Notices of the Royal Astronomical Society, 369, 216

\bibitem[{Alexander {et~al.}(2006{\natexlab{b}})Alexander, Clarke, \&
  Pringle}]{alexander2006photoevaporationII}
---. 2006{\natexlab{b}}, Monthly Notices of the Royal Astronomical Society,
  369, 229

\bibitem[{Andrews {et~al.}(2010)Andrews, Wilner, Hughes, Qi, \&
  Dullemond}]{andrews2010protoplanetary}
Andrews, S.~M., Wilner, D., Hughes, A., Qi, C., \& Dullemond, C. 2010, The
  Astrophysical Journal, 723, 1241

\bibitem[{Andrews {et~al.}(2018{\natexlab{a}})Andrews, Wilner, Macias,
  Carrasco-Gonzalez, \& Isella}]{andrews2018science}
Andrews, S.~M., Wilner, D.~J., Macias, E., Carrasco-Gonzalez, C., \& Isella, A.
  2018{\natexlab{a}}, arXiv preprint arXiv:1810.06598

\bibitem[{Andrews {et~al.}(2016)Andrews, Wilner, Zhu, Birnstiel, Carpenter,
  P{\'e}rez, Bai, {\"O}berg, Hughes, Isella, {et~al.}}]{andrews2016ringed}
Andrews, S.~M., Wilner, D.~J., Zhu, Z., {et~al.} 2016, The Astrophysical
  Journal Letters, 820, L40

\bibitem[{Andrews {et~al.}(2018{\natexlab{b}})Andrews, Huang, P{\'e}rez,
  Isella, Dullemond, Kurtovic, Guzm{\'a}n, Carpenter, Wilner, Zhang,
  {et~al.}}]{andrews2018disk}
Andrews, S.~M., Huang, J., P{\'e}rez, L.~M., {et~al.} 2018{\natexlab{b}}, The
  Astrophysical Journal Letters, 869, L41

\bibitem[{Ansdell {et~al.}(2016)Ansdell, Williams, van~der Marel, Carpenter,
  Guidi, Hogerheijde, Mathews, Manara, Miotello, Natta,
  {et~al.}}]{ansdell2016alma}
Ansdell, M., Williams, J.~P., van~der Marel, N., {et~al.} 2016, The
  Astrophysical Journal, 828, 46

\bibitem[{Armitage(2007)}]{armitage2007lecture}
Armitage, P.~J. 2007, arXiv preprint astro-ph/0701485

\bibitem[{Armitage(2011)}]{armitage2011dynamics}
---. 2011, Annual Review of Astronomy and Astrophysics, 49

\bibitem[{Bae \& Zhu(2018{\natexlab{a}})}]{bae2018planet1}
Bae, J., \& Zhu, Z. 2018{\natexlab{a}}, The Astrophysical Journal, 859

\bibitem[{Bae \& Zhu(2018{\natexlab{b}})}]{bae2018planet2}
---. 2018{\natexlab{b}}, The Astrophysical Journal, 859, 119

\bibitem[{Bae {et~al.}(2017)Bae, Zhu, \& Hartmann}]{bae2017formation}
Bae, J., Zhu, Z., \& Hartmann, L. 2017, The Astrophysical Journal, 850

\bibitem[{Bai(2013)}]{bai2013wind2}
Bai, X.-N. 2013, The Astrophysical Journal, 772, 96

\bibitem[{Bai(2014{\natexlab{a}})}]{bai2014hall1}
---. 2014{\natexlab{a}}, The Astrophysical Journal, 791, 137

\bibitem[{Bai(2014{\natexlab{b}})}]{bai2014hall2}
---. 2014{\natexlab{b}}, The Astrophysical Journal, 798, 84

\bibitem[{Bai \& Stone(2011)}]{bai2011ambipolar}
Bai, X.-N., \& Stone, J.~M. 2011, The Astrophysical Journal, 736, 144

\bibitem[{Bai \& Stone(2013)}]{bai2013wind1}
---. 2013, The Astrophysical Journal, 769, 76

\bibitem[{Bai {et~al.}(2016)Bai, Ye, Goodman, \& Yuan}]{bai2016magneto}
Bai, X.-N., Ye, J., Goodman, J., \& Yuan, F. 2016, The Astrophysical Journal,
  818, 152

\bibitem[{Balbus \& Hawley(1998)}]{balbus1998instability}
Balbus, S.~A., \& Hawley, J.~F. 1998, Reviews of modern physics, 70, 1

\bibitem[{Baruteau {et~al.}(2019)Baruteau, Barraza, P{\'e}rez, Casassus, Dong,
  Lyra, Marino, Christiaens, Zhu, Carmona, {et~al.}}]{baruteau2019dust}
Baruteau, C., Barraza, M., P{\'e}rez, S., {et~al.} 2019, Monthly Notices of the
  Royal Astronomical Society, 486, 304

\bibitem[{Beckwith {et~al.}(1990)Beckwith, Sargent, Chini, \&
  Guesten}]{beckwith1990survey}
Beckwith, S.~V., Sargent, A.~I., Chini, R.~S., \& Guesten, R. 1990, The
  Astronomical Journal, 99, 924

\bibitem[{Benisty {et~al.}(2015)Benisty, Juhasz, Boccaletti, Avenhaus, Milli,
  Thalmann, Dominik, Pinilla, Buenzli, Pohl, {et~al.}}]{benisty2015asymmetric}
Benisty, M., Juhasz, A., Boccaletti, A., {et~al.} 2015, Astronomy \&
  Astrophysics, 578, L6

\bibitem[{Benisty {et~al.}(2018)Benisty, Juh{\'a}sz, Facchini, Pinilla,
  de~Boer, P{\'e}rez, Keppler, Muro-Arena, Villenave, Andrews,
  {et~al.}}]{benisty2018shadows}
Benisty, M., Juh{\'a}sz, A., Facchini, S., {et~al.} 2018, Astronomy and
  Astrophysics, 619

\bibitem[{Birnstiel {et~al.}(2018)Birnstiel, Dullemond, Zhu, Andrews, Bai,
  Wilner, Carpenter, Huang, Isella, Benisty, {et~al.}}]{birnstiel2018disk}
Birnstiel, T., Dullemond, C.~P., Zhu, Z., {et~al.} 2018, The Astrophysical
  Journal Letters, 869, L45

\bibitem[{Blandford \& Payne(1982)}]{blandford1982hydromagnetic}
Blandford, R., \& Payne, D. 1982, Monthly Notices of the Royal Astronomical
  Society, 199, 883

\bibitem[{Boehler {et~al.}(2017)Boehler, Weaver, Isella, Ricci, Grady,
  Carpenter, \& Perez}]{boehler2017close}
Boehler, Y., Weaver, E., Isella, A., {et~al.} 2017, The Astrophysical Journal,
  840, 60

\bibitem[{Boehler {et~al.}(2018)Boehler, Ricci, Weaver, Isella, Benisty,
  Carpenter, Grady, Shen, Tang, \& Perez}]{boehler2018complex}
Boehler, Y., Ricci, L., Weaver, E., {et~al.} 2018, The Astrophysical Journal,
  853, 162

\bibitem[{Brogan {et~al.}(2015)Brogan, P{\'e}rez, Hunter, Dent, Hales, Hills,
  Corder, Fomalont, Vlahakis, Asaki, {et~al.}}]{almapartnetship2015}
Brogan, C., P{\'e}rez, L., Hunter, T., {et~al.} 2015, The Astrophysical Journal
  Letters, 808, L3

\bibitem[{Brown {et~al.}(2018)Brown, Vallenari, Prusti, De~Bruijne, Babusiaux,
  Bailer-Jones, Biermann, Evans, Eyer, Jansen, {et~al.}}]{brown2018gaia}
Brown, A., Vallenari, A., Prusti, T., {et~al.} 2018, Astronomy \& astrophysics,
  616, A1

\bibitem[{Bryden {et~al.}(1999)Bryden, Chen, Lin, Nelson, \&
  Papaloizou}]{bryden1999tidally}
Bryden, G., Chen, X., Lin, D., Nelson, R.~P., \& Papaloizou, J.~C. 1999, The
  Astrophysical Journal, 514, 344

\bibitem[{Bryden {et~al.}(2000)Bryden, R{\'o}{\.z}yczka, Lin, \&
  Bodenheimer}]{bryden2000interaction}
Bryden, G., R{\'o}{\.z}yczka, M., Lin, D., \& Bodenheimer, P. 2000, The
  Astrophysical Journal, 540, 1091

\bibitem[{Carrasco-Gonz{\'a}lez {et~al.}(2016)Carrasco-Gonz{\'a}lez, Henning,
  Chandler, Linz, P{\'e}rez, Rodr{\'\i}guez, Galv{\'a}n-Madrid, Anglada,
  Birnstiel, van Boekel, {et~al.}}]{carrasco2016vla}
Carrasco-Gonz{\'a}lez, C., Henning, T., Chandler, C.~J., {et~al.} 2016, The
  astrophysical journal letters, 821, L16

\bibitem[{Cazzoletti {et~al.}(2018)Cazzoletti, van Dishoeck, Pinilla, Tazzari,
  Facchini, van~der Marel, Benisty, Garufi, \&
  P{\'e}rez}]{cazzoletti2018evidence}
Cazzoletti, P., van Dishoeck, E., Pinilla, P., {et~al.} 2018, Astronomy \&
  Astrophysics, 619, A161

\bibitem[{Chiang \& Youdin(2010)}]{chiang2010forming}
Chiang, E., \& Youdin, A. 2010, Annual Review of Earth and Planetary Sciences,
  38, 493

\bibitem[{Demyk {et~al.}(2017{\natexlab{a}})Demyk, Meny, Lu, Papatheodorou,
  Toplis, Leroux, Depecker, Brubach, Roy, Nayral, {et~al.}}]{demyk2017low1}
Demyk, K., Meny, C., Lu, X.-H., {et~al.} 2017{\natexlab{a}}, Astronomy \&
  Astrophysics, 600, A123

\bibitem[{Demyk {et~al.}(2017{\natexlab{b}})Demyk, Meny, Leroux, Depecker,
  Brubach, Roy, Nayral, Ojo, \& Delpech}]{demyk2017low2}
Demyk, K., Meny, C., Leroux, H., {et~al.} 2017{\natexlab{b}}, Astronomy \&
  Astrophysics, 606, A50

\bibitem[{Dipierro {et~al.}(2015)Dipierro, Price, Laibe, Hirsh, Cerioli, \&
  Lodato}]{dipierro2015planet}
Dipierro, G., Price, D., Laibe, G., {et~al.} 2015, Monthly Notices of the Royal
  Astronomical Society: Letters, 453, L73

\bibitem[{Dong {et~al.}(2017)Dong, Li, Chiang, \& Li}]{dong2017multiple}
Dong, R., Li, S., Chiang, E., \& Li, H. 2017, The Astrophysical Journal, 843,
  127

\bibitem[{Dong {et~al.}(2018{\natexlab{a}})Dong, Li, Chiang, \&
  Li}]{dong2018multiple}
---. 2018{\natexlab{a}}, The Astrophysical Journal, 866, 110

\bibitem[{Dong {et~al.}(2011{\natexlab{a}})Dong, Rafikov, \&
  Stone}]{dong2011density2}
Dong, R., Rafikov, R.~R., \& Stone, J.~M. 2011{\natexlab{a}}, The Astrophysical
  Journal, 741, 57

\bibitem[{Dong {et~al.}(2011{\natexlab{b}})Dong, Rafikov, Stone, \&
  Petrovich}]{dong2011density1}
Dong, R., Rafikov, R.~R., Stone, J.~M., \& Petrovich, C. 2011{\natexlab{b}},
  The Astrophysical Journal, 741, 56

\bibitem[{Dong {et~al.}(2015)Dong, Zhu, \& Whitney}]{dong2015observational}
Dong, R., Zhu, Z., \& Whitney, B. 2015, The Astrophysical Journal, 809, 93

\bibitem[{Dong {et~al.}(2018{\natexlab{b}})Dong, Liu, Eisner, Andrews, Fung,
  Zhu, Chiang, Hashimoto, Liu, Casassus, {et~al.}}]{dong2018eccentric}
Dong, R., Liu, S.-y., Eisner, J., {et~al.} 2018{\natexlab{b}}, The
  Astrophysical Journal, 860, 124

\bibitem[{Draine(2006)}]{draine2006submillimeter}
Draine, B.~T. 2006, The Astrophysical Journal, 636, 1114

\bibitem[{Dra{\.z}kowska {et~al.}(2019)Dra{\.z}kowska, Li, Birnstiel, Stammler,
  \& Li}]{drazkowska2019including}
Dra{\.z}kowska, J., Li, S., Birnstiel, T., Stammler, S.~M., \& Li, H. 2019, The
  Astrophysical Journal, 885, 91

\bibitem[{Dullemond(2012)}]{dullemond2012radmc}
Dullemond, C. 2012, Astrophysics Source Code Library

\bibitem[{Dullemond \& Penzlin(2018)}]{dullemond2018dust}
Dullemond, C., \& Penzlin, A. 2018, Astronomy \& Astrophysics, 609, A50

\bibitem[{Dullemond {et~al.}(2018)Dullemond, Birnstiel, Huang, Kurtovic,
  Andrews, Guzm{\'a}n, P{\'e}rez, Isella, Zhu, Benisty,
  {et~al.}}]{dullemond2018disk}
Dullemond, C.~P., Birnstiel, T., Huang, J., {et~al.} 2018, The Astrophysical
  Journal Letters, 869, L46

\bibitem[{Fedele {et~al.}(2018)Fedele, Tazzari, Booth, Testi, Clarke, Pascucci,
  Kospal, Semenov, Bruderer, Henning, {et~al.}}]{fedele2018alma}
Fedele, D., Tazzari, M., Booth, R., {et~al.} 2018, Astronomy \& Astrophysics,
  610, A24

\bibitem[{Flock {et~al.}(2015)Flock, Ruge, Dzyurkevich, Henning, Klahr, \&
  Wolf}]{flock2015gaps}
Flock, M., Ruge, J., Dzyurkevich, N., {et~al.} 2015, Astronomy \& Astrophysics,
  574, A68

\bibitem[{Frank {et~al.}(2002)Frank, King, \& Raine}]{frank2002accretion}
Frank, J., King, A., \& Raine, D. 2002, Accretion power in astrophysics
  (Cambridge university press), 88--89

\bibitem[{Fu {et~al.}(2014{\natexlab{a}})Fu, Li, Lubow, \& Li}]{fu2014long}
Fu, W., Li, H., Lubow, S., \& Li, S. 2014{\natexlab{a}}, The Astrophysical
  Journal Letters, 788, L41

\bibitem[{Fu {et~al.}(2014{\natexlab{b}})Fu, Li, Lubow, Li, \&
  Liang}]{fu2014effects}
Fu, W., Li, H., Lubow, S., Li, S., \& Liang, E. 2014{\natexlab{b}}, The
  Astrophysical Journal Letters, 795, L39

\bibitem[{Fujiwara {et~al.}(2006)Fujiwara, Honda, Kataza, Yamashita, Onaka,
  Fukagawa, Okamoto, Miyata, Sako, Fujiyoshi,
  {et~al.}}]{fujiwara2006asymmetric}
Fujiwara, H., Honda, M., Kataza, H., {et~al.} 2006, The Astrophysical Journal
  Letters, 644, L133

\bibitem[{{Fung} {et~al.}(2014){Fung}, {Shi}, \& {Chiang}}]{fung2014empty}
{Fung}, J., {Shi}, J.-M., \& {Chiang}, E. 2014, The Astrophysical Journal, 782,
  88

\bibitem[{Gammie(1996)}]{gammie1996layered}
Gammie, C.~F. 1996, The Astrophysical Journal, 457, 355

\bibitem[{Goldreich \& Lynden-Bell(1965)}]{goldreich1965gravitational}
Goldreich, P., \& Lynden-Bell, D. 1965, Monthly Notices of the Royal
  Astronomical Society, 130, 97

\bibitem[{Goldreich \& Tremaine(1978)}]{goldreich1978excitation}
Goldreich, P., \& Tremaine, S. 1978, The Astrophysical Journal, 222, 850

\bibitem[{Goldreich \& Tremaine(1979)}]{goldreich1979excitation}
---. 1979, Astrophysical Journal, 233, 857

\bibitem[{Goldreich \& Tremaine(1980)}]{goldreich1980disk}
---. 1980, Astrophysical Journal, 241, 425

\bibitem[{Goodman \& Pindor(2000)}]{goodman2000secular}
Goodman, J., \& Pindor, B. 2000, Icarus, 148, 537

\bibitem[{Goodman \& Rafikov(2001)}]{goodman2001planetary}
Goodman, J., \& Rafikov, R. 2001, The Astrophysical Journal, 552, 793

\bibitem[{Grady {et~al.}(2012)Grady, Muto, Hashimoto, Fukagawa, Currie, Biller,
  Thalmann, Sitko, Russell, Wisniewski, {et~al.}}]{grady2012spiral}
Grady, C., Muto, T., Hashimoto, J., {et~al.} 2012, The Astrophysical Journal,
  762, 48

\bibitem[{Guzm{\'a}n {et~al.}(2018)Guzm{\'a}n, Huang, Andrews, Isella,
  P{\'e}rez, Carpenter, Dullemond, Ricci, Birnstiel, Zhang,
  {et~al.}}]{guzman2018disk}
Guzm{\'a}n, V.~V., Huang, J., Andrews, S.~M., {et~al.} 2018, The Astrophysical
  Journal Letters, 869, L48

\bibitem[{Huang {et~al.}(2018{\natexlab{a}})Huang, Andrews, Cleeves, {\"O}berg,
  Wilner, Bai, Birnstiel, Carpenter, Hughes, Isella, {et~al.}}]{huang2018co}
Huang, J., Andrews, S.~M., Cleeves, L.~I., {et~al.} 2018{\natexlab{a}}, The
  Astrophysical Journal, 852, 122

\bibitem[{Huang {et~al.}(2018{\natexlab{b}})Huang, Andrews, Dullemond, Isella,
  P{\'e}rez, Guzm{\'a}n, {\"O}berg, Zhu, Zhang, Bai, {et~al.}}]{huang2018disk2}
Huang, J., Andrews, S.~M., Dullemond, C.~P., {et~al.} 2018{\natexlab{b}}, The
  Astrophysical Journal Letters, 869, L42

\bibitem[{Huang {et~al.}(2018{\natexlab{c}})Huang, Andrews, P{\'e}rez, Zhu,
  Dullemond, Isella, Benisty, Bai, Birnstiel, Carpenter,
  {et~al.}}]{huang2018disk3}
Huang, J., Andrews, S.~M., P{\'e}rez, L.~M., {et~al.} 2018{\natexlab{c}}, The
  Astrophysical Journal Letters, 869, L43

\bibitem[{Huang {et~al.}(2019)Huang, Dong, Li, Li, \&
  Ji}]{huang2019observability}
Huang, P., Dong, R., Li, H., Li, S., \& Ji, J. 2019, The Astrophysical Journal
  Letters, 883, L39

\bibitem[{Huang {et~al.}(2018{\natexlab{d}})Huang, Isella, Li, Li, \&
  Ji}]{huang2018identifying}
Huang, P., Isella, A., Li, H., Li, S., \& Ji, J. 2018{\natexlab{d}}, The
  Astrophysical Journal, 867, 3

\bibitem[{Isella {et~al.}(2009)Isella, Carpenter, \&
  Sargent}]{isella2009structure}
Isella, A., Carpenter, J.~M., \& Sargent, A.~I. 2009, The Astrophysical
  Journal, 701, 260

\bibitem[{Isella {et~al.}(2010)Isella, Natta, Wilner, Carpenter, \&
  Testi}]{isella2010millimeter}
Isella, A., Natta, A., Wilner, D., Carpenter, J.~M., \& Testi, L. 2010, The
  Astrophysical Journal, 725, 1735

\bibitem[{Isella {et~al.}(2013)Isella, P{\'e}rez, Carpenter, Ricci, Andrews, \&
  Rosenfeld}]{isella2013azimuthal}
Isella, A., P{\'e}rez, L.~M., Carpenter, J.~M., {et~al.} 2013, The
  Astrophysical Journal, 775, 30

\bibitem[{Isella {et~al.}(2007)Isella, Testi, Natta, Neri, Wilner, \&
  Qi}]{isella2007millimeter}
Isella, A., Testi, L., Natta, A., {et~al.} 2007, Astronomy \& Astrophysics,
  469, 213

\bibitem[{Isella {et~al.}(2016)Isella, Guidi, Testi, Liu, Li, Li, Weaver,
  Boehler, Carperter, De~Gregorio-Monsalvo, {et~al.}}]{isella2016ringed}
Isella, A., Guidi, G., Testi, L., {et~al.} 2016, Physical Review Letters, 117,
  251101

\bibitem[{Isella {et~al.}(2018)Isella, Huang, Andrews, Dullemond, Birnstiel,
  Zhang, Zhu, Guzm{\'a}n, P{\'e}rez, Bai, {et~al.}}]{isella2018disk}
Isella, A., Huang, J., Andrews, S.~M., {et~al.} 2018, The Astrophysical Journal
  Letters, 869, L49

\bibitem[{Jin {et~al.}(2019)Jin, Isella, Huang, Li, Li, \& Ji}]{jin2019new}
Jin, S., Isella, A., Huang, P., {et~al.} 2019, The Astrophysical Journal, 881,
  108

\bibitem[{Jin {et~al.}(2016)Jin, Li, Isella, Li, \& Ji}]{jin2016modeling}
Jin, S., Li, S., Isella, A., Li, H., \& Ji, J. 2016, The Astrophysical Journal,
  818, 76

\bibitem[{Johansen \& Youdin(2007)}]{johansen2007protoplanetary}
Johansen, A., \& Youdin, A. 2007, The Astrophysical Journal, 662, 627

\bibitem[{Jones {et~al.}(2016)Jones, Oliphant, Peterson,
  {et~al.}}]{jones2016scipy}
Jones, E., Oliphant, T., Peterson, P., {et~al.} 2016, SciPy: Open source
  scientific tools for Python, 2001

\bibitem[{Kanagawa {et~al.}(2018)Kanagawa, Muto, Okuzumi, Tanigawa, Taki, \&
  Shibaike}]{kanagawa2018impacts}
Kanagawa, K.~D., Muto, T., Okuzumi, S., {et~al.} 2018, The Astrophysical
  Journal, 868, 48

\bibitem[{Kataoka {et~al.}(2015)Kataoka, Muto, Momose, Tsukagoshi, Fukagawa,
  Shibai, Hanawa, Murakawa, \& Dullemond}]{kataoka2015millimeter}
Kataoka, A., Muto, T., Momose, M., {et~al.} 2015, The Astrophysical Journal,
  809, 78

\bibitem[{Li {et~al.}(2001)Li, Colgate, Wendroff, \& Liska}]{li2001rossby}
Li, H., Colgate, S., Wendroff, B., \& Liska, R. 2001, The Astrophysical
  Journal, 551, 874

\bibitem[{Li {et~al.}(2000)Li, Finn, Lovelace, \& Colgate}]{li2000rossby}
Li, H., Finn, J., Lovelace, R., \& Colgate, S. 2000, The Astrophysical Journal,
  533, 1023

\bibitem[{Li {et~al.}(2005)Li, Li, Koller, Wendroff, Liska, Orban, Liang, \&
  Lin}]{li2005potential}
Li, H., Li, S., Koller, J., {et~al.} 2005, The Astrophysical Journal, 624, 1003

\bibitem[{Li {et~al.}(2008)Li, Lubow, Li, \& Lin}]{li2008type}
Li, H., Lubow, S., Li, S., \& Lin, D.~N. 2008, The Astrophysical Journal
  Letters, 690, L52

\bibitem[{Liu(2019)}]{liu2019anomalously}
Liu, H.~B. 2019, The Astrophysical Journal Letters, 877, L22

\bibitem[{Liu {et~al.}(2018{\natexlab{a}})Liu, Ricci, Isella, Li, \&
  Li}]{liu2018ngvla}
Liu, S., Ricci, L., Isella, A., Li, H., \& Li, S. 2018{\natexlab{a}}, in
  American Astronomical Society Meeting Abstracts, Vol. 231

\bibitem[{Liu {et~al.}(2018{\natexlab{b}})Liu, Jin, Li, Isella, \&
  Li}]{liu2018new}
Liu, S.-F., Jin, S., Li, S., Isella, A., \& Li, H. 2018{\natexlab{b}}, The
  Astrophysical Journal, 857, 87

\bibitem[{Loinard {et~al.}(2007)Loinard, Torres, Mioduszewski, Rodr{\'\i}guez,
  Gonz{\'a}lez-L{\'o}pezlira, Lachaume, V{\'a}zquez, \&
  Gonz{\'a}lez}]{loinard2007vlba}
Loinard, L., Torres, R.~M., Mioduszewski, A.~J., {et~al.} 2007, The
  Astrophysical Journal, 671, 546

\bibitem[{Long {et~al.}(2018)Long, Pinilla, Herczeg, Harsono, Dipierro,
  Pascucci, Hendler, Tazzari, Ragusa, Salyk, {et~al.}}]{long2018gaps}
Long, F., Pinilla, P., Herczeg, G.~J., {et~al.} 2018, The Astrophysical
  Journal, 869, 17

\bibitem[{Lovelace {et~al.}(1999)Lovelace, Li, Colgate, \&
  Nelson}]{lovelace1999rossby}
Lovelace, R., Li, H., Colgate, S., \& Nelson, A. 1999, The Astrophysical
  Journal, 513, 805

\bibitem[{Lovelace \& Romanova(2014)}]{lovelace2014rossby}
Lovelace, R., \& Romanova, M. 2014, Fluid Dynamics Research, 46, 041401

\bibitem[{Lyra {et~al.}(2009)Lyra, Johansen, Zsom, Klahr, \&
  Piskunov}]{lyra2009planet}
Lyra, W., Johansen, A., Zsom, A., Klahr, H., \& Piskunov, N. 2009, Astronomy \&
  Astrophysics, 497, 869

\bibitem[{Marino {et~al.}(2015)Marino, Casassus, Perez, Lyra, Roman, Avenhaus,
  Wright, \& Maddison}]{marino2015compact}
Marino, S., Casassus, S., Perez, S., {et~al.} 2015, The Astrophysical Journal,
  813, 76

\bibitem[{McMullin {et~al.}(2007)McMullin, Waters, Schiebel, Young, \&
  Golap}]{mcmullin2007casa}
McMullin, J., Waters, B., Schiebel, D., Young, W., \& Golap, K. 2007, in
  Astronomical data analysis software and systems XVI, Vol. 376, 127

\bibitem[{Meeus {et~al.}(2012)Meeus, Montesinos, Mendigut{\'\i}a, Kamp, Thi,
  Eiroa, Grady, Mathews, Sandell, Martin-Za{\"\i}di,
  {et~al.}}]{meeus2012observations}
Meeus, G., Montesinos, B., Mendigut{\'\i}a, I., {et~al.} 2012, Astronomy \&
  Astrophysics, 544, A78

\bibitem[{Miranda {et~al.}(2017)Miranda, Li, Li, \& Jin}]{miranda2017long}
Miranda, R., Li, H., Li, S., \& Jin, S. 2017, The Astrophysical Journal, 835,
  118

\bibitem[{Miranda \& Rafikov(2019{\natexlab{a}})}]{miranda2019gaps}
Miranda, R., \& Rafikov, R.~R. 2019{\natexlab{a}}, The Astrophysical Journal
  Letters, 878, L9

\bibitem[{Miranda \& Rafikov(2019{\natexlab{b}})}]{miranda2019multiple}
---. 2019{\natexlab{b}}, The Astrophysical Journal, 875, 37

\bibitem[{Ono {et~al.}(2016)Ono, Muto, Takeuchi, \& Nomura}]{ono2016parametric}
Ono, T., Muto, T., Takeuchi, T., \& Nomura, H. 2016, The Astrophysical Journal,
  823, 84

\bibitem[{Ono {et~al.}(2018)Ono, Muto, Tomida, \& Zhu}]{ono2018parametric}
Ono, T., Muto, T., Tomida, K., \& Zhu, Z. 2018, The Astrophysical Journal, 864,
  70

\bibitem[{Ou {et~al.}(2007)Ou, Ji, Liu, \& Peng}]{ou2007disk}
Ou, S., Ji, J., Liu, L., \& Peng, X. 2007, The Astrophysical Journal, 667, 1220

\bibitem[{P{\'e}rez {et~al.}(2018{\natexlab{a}})P{\'e}rez, Andrews, Isella, \&
  Dullemond}]{perez2018first}
P{\'e}rez, L.~M., Andrews, S., Isella, A., \& Dullemond, K. 2018{\natexlab{a}},
  in Diversis Mundi: The Solar System in an Exoplanetary Context

\bibitem[{P{\'e}rez {et~al.}(2014)P{\'e}rez, Isella, Carpenter, \&
  Chandler}]{perez2014large}
P{\'e}rez, L.~M., Isella, A., Carpenter, J.~M., \& Chandler, C.~J. 2014, The
  Astrophysical Journal Letters, 783, L13

\bibitem[{P{\'e}rez {et~al.}(2018{\natexlab{b}})P{\'e}rez, Benisty, Andrews,
  Isella, Dullemond, Huang, Kurtovic, Guzm{\'a}n, Zhu, Birnstiel,
  {et~al.}}]{perez2018disk}
P{\'e}rez, L.~M., Benisty, M., Andrews, S.~M., {et~al.} 2018{\natexlab{b}}, The
  Astrophysical Journal Letters, 869, L50

\bibitem[{Pierens {et~al.}(2019)Pierens, Lin, \& Raymond}]{pierens2019vortex}
Pierens, A., Lin, M.-K., \& Raymond, S.~N. 2019, Monthly Notices of the Royal
  Astronomical Society, 488, 645

\bibitem[{Pinte {et~al.}(2015)Pinte, Dent, Menard, Hales, Hill, Cortes, \&
  de~Gregorio-Monsalvo}]{pinte2015dust}
Pinte, C., Dent, W.~R., Menard, F., {et~al.} 2015, The Astrophysical Journal,
  816, 25

\bibitem[{Pollack {et~al.}(1994)Pollack, Hollenbach, Beckwith, Simonelli,
  Roush, \& Fong}]{pollack1994composition}
Pollack, J.~B., Hollenbach, D., Beckwith, S., {et~al.} 1994, The Astrophysical
  Journal, 421, 615

\bibitem[{Pringle(1981)}]{pringle1981accretion}
Pringle, J. 1981, Annual review of astronomy and astrophysics, 19, 137

\bibitem[{Rafikov(2002)}]{rafikov2002planet}
Rafikov, R.~R. 2002, The Astrophysical Journal, 572, 566

\bibitem[{Reg{\'a}ly {et~al.}(2011)Reg{\'a}ly, Juh{\'a}sz, S{\'a}ndor, \&
  Dullemond}]{regaly2011possible}
Reg{\'a}ly, Z., Juh{\'a}sz, A., S{\'a}ndor, Z., \& Dullemond, C. 2011, Monthly
  Notices of the Royal Astronomical Society, 419, 1701

\bibitem[{Reg{\'a}ly {et~al.}(2013)Reg{\'a}ly, S{\'a}ndor, Csom{\'o}s, \&
  Ataiee}]{regaly2013trapping}
Reg{\'a}ly, Z., S{\'a}ndor, Z., Csom{\'o}s, P., \& Ataiee, S. 2013, Monthly
  Notices of the Royal Astronomical Society, 433, 2626

\bibitem[{Robitaille {et~al.}(2013)Robitaille, Tollerud, Greenfield,
  Droettboom, Bray, Aldcroft, Davis, Ginsburg, Price-Whelan, Kerzendorf,
  {et~al.}}]{robitaille2013astropy}
Robitaille, T.~P., Tollerud, E.~J., Greenfield, P., {et~al.} 2013, Astronomy \&
  Astrophysics, 558, A33

\bibitem[{Semenov {et~al.}(2003)Semenov, Henning, Helling, Ilgner, \&
  Sedlmayr}]{semenov2003rosseland}
Semenov, D., Henning, T., Helling, C., Ilgner, M., \& Sedlmayr, E. 2003,
  Astronomy \& Astrophysics, 410, 611

\bibitem[{Shakura \& Sunyaev(1973)}]{shakura1973black}
Shakura, N.~I., \& Sunyaev, R.~A. 1973, Astronomy and Astrophysics, 24, 337

\bibitem[{Takeuchi \& Lin(2002)}]{takeuchi2002radial}
Takeuchi, T., \& Lin, D. 2002, The Astrophysical Journal, 581, 1344

\bibitem[{Taki {et~al.}(2016)Taki, Fujimoto, \& Ida}]{taki2016dust}
Taki, T., Fujimoto, M., \& Ida, S. 2016, Astronomy \& Astrophysics, 591, A86

\bibitem[{Testi {et~al.}(2014)Testi, Birnstiel, Ricci, Andrews, Blum,
  Carpenter, Dominik, Isella, Natta, Williams, {et~al.}}]{testi2014dust}
Testi, L., Birnstiel, T., Ricci, L., {et~al.} 2014, Protostars and Planets VI,
  339

\bibitem[{Toomre(1964)}]{toomre1964gravitational}
Toomre, A. 1964, The Astrophysical Journal, 139, 1217

\bibitem[{Uyama {et~al.}(2018)Uyama, Hashimoto, Muto, Akiyama, Dong, De~Leon,
  Sakon, Kudo, Kusakabe, Kuzuhara, {et~al.}}]{uyama2018subaru}
Uyama, T., Hashimoto, J., Muto, T., {et~al.} 2018, The Astronomical Journal,
  156, 63

\bibitem[{van~der Marel {et~al.}(2016)van~der Marel, Cazzoletti, Pinilla, \&
  Garufi}]{van2016vortices}
van~der Marel, N., Cazzoletti, P., Pinilla, P., \& Garufi, A. 2016, The
  Astrophysical Journal, 832, 178

\bibitem[{van~der Marel {et~al.}(2018)van~der Marel, Williams, \&
  Bruderer}]{van2018rings}
van~der Marel, N., Williams, J.~P., \& Bruderer, S. 2018, The Astrophysical
  Journal Letters, 867, L14

\bibitem[{van~der Marel {et~al.}(2013)van~der Marel, Van~Dishoeck, Bruderer,
  Birnstiel, Pinilla, Dullemond, van Kempen, Schmalzl, Brown, Herczeg,
  {et~al.}}]{van2013major}
van~der Marel, N., Van~Dishoeck, E.~F., Bruderer, S., {et~al.} 2013, Science,
  340, 1199

\bibitem[{Walt {et~al.}(2011)Walt, Colbert, \& Varoquaux}]{walt2011numpy}
Walt, S. v.~d., Colbert, S.~C., \& Varoquaux, G. 2011, Computing in Science \&
  Engineering, 13, 22

\bibitem[{Weidenschilling(1977)}]{weidenschilling1977aerodynamics}
Weidenschilling, S. 1977, Monthly Notices of the Royal Astronomical Society,
  180, 57

\bibitem[{Weingartner \& Draine(2001)}]{weingartner2001dust}
Weingartner, J.~C., \& Draine, B. 2001, The Astrophysical Journal, 548, 296

\bibitem[{Yang \& Zhu(2019)}]{yang2019morphological}
Yang, C.-C., \& Zhu, Z. 2019, Monthly Notices of the Royal Astronomical
  Society, 491, 4702

\bibitem[{Yen {et~al.}(2016)Yen, Liu, Gu, Hirano, Lee, Puspitaningrum, \&
  Takakuwa}]{yen2016gas}
Yen, H.-W., Liu, H.~B., Gu, P.-G., {et~al.} 2016, The Astrophysical Journal
  Letters, 820, L25

\bibitem[{Youdin \& Johansen(2007)}]{youdin2007protoplanetary}
Youdin, A., \& Johansen, A. 2007, The Astrophysical Journal, 662, 613

\bibitem[{Youdin \& Goodman(2005)}]{youdin2005streaming}
Youdin, A.~N., \& Goodman, J. 2005, The Astrophysical Journal, 620, 459

\bibitem[{Zhang {et~al.}(2016)Zhang, Bergin, Blake, Cleeves, Hogerheijde,
  Salinas, \& Schwarz}]{zhang2016commonality}
Zhang, K., Bergin, E.~A., Blake, G.~A., {et~al.} 2016, The Astrophysical
  Journal Letters, 818, L16

\bibitem[{Zhang {et~al.}(2015)Zhang, Blake, \& Bergin}]{zhang2015evidence}
Zhang, K., Blake, G.~A., \& Bergin, E.~A. 2015, The Astrophysical Journal
  Letters, 806, L7

\bibitem[{Zhang {et~al.}(2018)Zhang, Zhu, Huang, Guzm{\'a}n, Andrews,
  Birnstiel, Dullemond, Carpenter, Isella, P{\'e}rez, {et~al.}}]{zhang2018disk}
Zhang, S., Zhu, Z., Huang, J., {et~al.} 2018, The Astrophysical Journal
  Letters, 869, L47

\bibitem[{Zhu \& Stone(2014)}]{zhu2014dust}
Zhu, Z., \& Stone, J.~M. 2014, The Astrophysical Journal, 795, 53

\bibitem[{Zhu {et~al.}(2014)Zhu, Stone, Rafikov, \& Bai}]{zhu2014particle}
Zhu, Z., Stone, J.~M., Rafikov, R.~R., \& Bai, X.-N. 2014, The Astrophysical
  Journal, 785, 122

\bibitem[{Zhu {et~al.}(2019)Zhu, Zhang, Jiang, Kataoka, Birnstiel, Dullemond,
  Andrews, Huang, P{\'e}rez, Carpenter, {et~al.}}]{zhu2019one}
Zhu, Z., Zhang, S., Jiang, Y.-F., {et~al.} 2019, The Astrophysical Journal
  Letters, 877, L18

\bibitem[{Zubko {et~al.}(1998)Zubko, Krelowski, \& Wegner}]{zubko1998size}
Zubko, V., Krelowski, J., \& Wegner, W. 1998, Monthly Notices of the Royal
  Astronomical Society, 294, 548

\end{thebibliography}

\end{document}